\documentclass[twocolumn,aps,prc,superscriptaddress,showpacs,nofootinbib,floatfix]{revtex4}

\usepackage{graphicx}	

\usepackage{dcolumn}	
\usepackage{multirow}	
\newcommand{\mean}[1]{$\left\langle #1 \right\rangle$}

\usepackage{xspace}

\newcommand{\pt}{\mbox{$p_T$}\xspace}

\newcommand{\raa}{\mbox{$R_{\rm AA}$}\xspace}
\newcommand{\df}{\mbox{$\Delta\phi$}\xspace}
\newcommand{\raadf}{\mbox{$R_{\rm AA}(\Delta\phi)$}\xspace}
\newcommand{\taa}{\mbox{$T_{\rm AA}$}\xspace}
\newcommand{\TAB}{\mbox{$T_{\rm AB}$}\xspace}

\newcommand{\Npart}{\mbox{$N_{\rm part}$}\xspace}
\newcommand{\Ncoll}{\mbox{$N_{\rm coll}$}\xspace}

\newcommand{\sqsn}{\mbox{$\sqrt{s_{_{NN}}}$}\xspace}

\newcommand{\auau}{\mbox{Au$+$Au}\xspace}
\newcommand{\pbpb}{\mbox{Pb$+$Pb}\xspace}
\newcommand{\dau}{\mbox{$d$Au}\xspace}

\newcommand{\pp}{\mbox{$p$$+$$p$}\xspace}
\newcommand{\piz}{\mbox{$\pi^0$}\xspace}
\newcommand{\gevc}{\mbox{GeV/$c$}\xspace}
\newcommand{\mgg}{\mbox{$m_{\gamma\gamma}$}\xspace}
\newcommand{\sloss}{\mbox{$S_{\rm loss}$}\xspace}
\def\fig#1{Figure~\ref{#1}}

\begin{document}


\title{Neutral pion production with respect to centrality and reaction
plane in Au$+$Au collisions at $\sqrt{s_{_{NN}}}$=200~GeV} 

\newcommand{\abilene}{Abilene Christian University, Abilene, Texas 79699, USA}
\newcommand{\augie}{Department of Physics, Augustana College, Sioux Falls, South Dakota 57197, USA}
\newcommand{\banaras}{Department of Physics, Banaras Hindu University, Varanasi 221005, India}
\newcommand{\barc}{Bhabha Atomic Research Centre, Bombay 400 085, India}
\newcommand{\bnlcoll}{Collider-Accelerator Department, Brookhaven National Laboratory, Upton, New York 11973-5000, USA}
\newcommand{\bnlphys}{Physics Department, Brookhaven National Laboratory, Upton, New York 11973-5000, USA}
\newcommand{\caucr}{University of California - Riverside, Riverside, California 92521, USA}
\newcommand{\charlesczech}{Charles University, Ovocn\'{y} trh 5, Praha 1, 116 36, Prague, Czech Republic}
\newcommand{\chonbuk}{Chonbuk National University, Jeonju, 561-756, Korea}
\newcommand{\ciae}{Science and Technology on Nuclear Data Laboratory, China Institute of Atomic Energy, Beijing 102413, P.~R.~China}
\newcommand{\cns}{Center for Nuclear Study, Graduate School of Science, University of Tokyo, 7-3-1 Hongo, Bunkyo, Tokyo 113-0033, Japan}
\newcommand{\colorado}{University of Colorado, Boulder, Colorado 80309, USA}
\newcommand{\columbia}{Columbia University, New York, New York 10027 and Nevis Laboratories, Irvington, New York 10533, USA}
\newcommand{\czechtech}{Czech Technical University, Zikova 4, 166 36 Prague 6, Czech Republic}
\newcommand{\dapnia}{Dapnia, CEA Saclay, F-91191, Gif-sur-Yvette, France}
\newcommand{\debrecen}{Debrecen University, H-4010 Debrecen, Egyetem t{\'e}r 1, Hungary}
\newcommand{\elte}{ELTE, E{\"o}tv{\"o}s Lor{\'a}nd University, H - 1117 Budapest, P{\'a}zm{\'a}ny P. s. 1/A, Hungary}
\newcommand{\ewha}{Ewha Womans University, Seoul 120-750, Korea}
\newcommand{\fit}{Florida Institute of Technology, Melbourne, Florida 32901, USA}
\newcommand{\fsu}{Florida State University, Tallahassee, Florida 32306, USA}
\newcommand{\gsu}{Georgia State University, Atlanta, Georgia 30303, USA}
\newcommand{\hiroshima}{Hiroshima University, Kagamiyama, Higashi-Hiroshima 739-8526, Japan}
\newcommand{\ihepprot}{IHEP Protvino, State Research Center of Russian Federation, Institute for High Energy Physics, Protvino, 142281, Russia}
\newcommand{\illuiuc}{University of Illinois at Urbana-Champaign, Urbana, Illinois 61801, USA}
\newcommand{\inrras}{Institute for Nuclear Research of the Russian Academy of Sciences, prospekt 60-letiya Oktyabrya 7a, Moscow 117312, Russia}
\newcommand{\instpasczech}{Institute of Physics, Academy of Sciences of the Czech Republic, Na Slovance 2, 182 21 Prague 8, Czech Republic}
\newcommand{\isu}{Iowa State University, Ames, Iowa 50011, USA}
\newcommand{\jaea}{Advanced Science Research Center, Japan Atomic Energy Agency, 2-4 Shirakata Shirane, Tokai-mura, Naka-gun, Ibaraki-ken 319-1195, Japan}
\newcommand{\jinrdubna}{Joint Institute for Nuclear Research, 141980 Dubna, Moscow Region, Russia}
\newcommand{\jyvaskyla}{Helsinki Institute of Physics and University of Jyv{\"a}skyl{\"a}, P.O.Box 35, FI-40014 Jyv{\"a}skyl{\"a}, Finland}
\newcommand{\kek}{KEK, High Energy Accelerator Research Organization, Tsukuba, Ibaraki 305-0801, Japan}
\newcommand{\korea}{Korea University, Seoul, 136-701, Korea}
\newcommand{\kurchatov}{Russian Research Center ``Kurchatov Institute", Moscow, 123098 Russia}
\newcommand{\kyoto}{Kyoto University, Kyoto 606-8502, Japan}
\newcommand{\labllr}{Laboratoire Leprince-Ringuet, Ecole Polytechnique, CNRS-IN2P3, Route de Saclay, F-91128, Palaiseau, France}
\newcommand{\lahorelums}{Physics Department, Lahore University of Management Sciences, Lahore, Pakistan}
\newcommand{\lawllnl}{Lawrence Livermore National Laboratory, Livermore, California 94550, USA}
\newcommand{\losalamos}{Los Alamos National Laboratory, Los Alamos, New Mexico 87545, USA}
\newcommand{\lpc}{LPC, Universit{\'e} Blaise Pascal, CNRS-IN2P3, Clermont-Fd, 63177 Aubiere Cedex, France}
\newcommand{\lund}{Department of Physics, Lund University, Box 118, SE-221 00 Lund, Sweden}
\newcommand{\maryland}{University of Maryland, College Park, Maryland 20742, USA}
\newcommand{\mass}{Department of Physics, University of Massachusetts, Amherst, Massachusetts 01003-9337, USA }
\newcommand{\michigan}{Department of Physics, University of Michigan, Ann Arbor, Michigan 48109-1040, USA}
\newcommand{\muenster}{Institut fur Kernphysik, University of Muenster, D-48149 Muenster, Germany}
\newcommand{\muhlenberg}{Muhlenberg College, Allentown, Pennsylvania 18104-5586, USA}
\newcommand{\myongji}{Myongji University, Yongin, Kyonggido 449-728, Korea}
\newcommand{\nagasaki}{Nagasaki Institute of Applied Science, Nagasaki-shi, Nagasaki 851-0193, Japan}
\newcommand{\newmex}{University of New Mexico, Albuquerque, New Mexico 87131, USA }
\newcommand{\nmsu}{New Mexico State University, Las Cruces, New Mexico 88003, USA}
\newcommand{\ohio}{Department of Physics and Astronomy, Ohio University, Athens, Ohio 45701, USA}
\newcommand{\ornl}{Oak Ridge National Laboratory, Oak Ridge, Tennessee 37831, USA}
\newcommand{\orsay}{IPN-Orsay, Universite Paris Sud, CNRS-IN2P3, BP1, F-91406, Orsay, France}
\newcommand{\peking}{Peking University, Beijing 100871, P.~R.~China}
\newcommand{\pnpi}{PNPI, Petersburg Nuclear Physics Institute, Gatchina, Leningrad region, 188300, Russia}
\newcommand{\riken}{RIKEN Nishina Center for Accelerator-Based Science, Wako, Saitama 351-0198, Japan}
\newcommand{\rikjrbrc}{RIKEN BNL Research Center, Brookhaven National Laboratory, Upton, New York 11973-5000, USA}
\newcommand{\rikkyo}{Physics Department, Rikkyo University, 3-34-1 Nishi-Ikebukuro, Toshima, Tokyo 171-8501, Japan}
\newcommand{\saispbstu}{Saint Petersburg State Polytechnic University, St. Petersburg, 195251 Russia}
\newcommand{\saopaulo}{Universidade de S{\~a}o Paulo, Instituto de F\'{\i}sica, Caixa Postal 66318, S{\~a}o Paulo CEP05315-970, Brazil}
\newcommand{\seoulnat}{Seoul National University, Seoul, Korea}
\newcommand{\stonybrkc}{Chemistry Department, Stony Brook University, SUNY, Stony Brook, New York 11794-3400, USA}
\newcommand{\stonycrkp}{Department of Physics and Astronomy, Stony Brook University, SUNY, Stony Brook, New York 11794-3400, USA}
\newcommand{\tenn}{University of Tennessee, Knoxville, Tennessee 37996, USA}
\newcommand{\titech}{Department of Physics, Tokyo Institute of Technology, Oh-okayama, Meguro, Tokyo 152-8551, Japan}
\newcommand{\tsukuba}{Institute of Physics, University of Tsukuba, Tsukuba, Ibaraki 305, Japan}
\newcommand{\vandy}{Vanderbilt University, Nashville, Tennessee 37235, USA}
\newcommand{\waseda}{Waseda University, Advanced Research Institute for Science and Engineering, 17 Kikui-cho, Shinjuku-ku, Tokyo 162-0044, Japan}
\newcommand{\weizmann}{Weizmann Institute, Rehovot 76100, Israel}
\newcommand{\wigner}{Institute for Particle and Nuclear Physics, Wigner Research Centre for Physics, Hungarian Academy of Sciences (Wigner RCP, RMKI) H-1525 Budapest 114, POBox 49, Budapest, Hungary}
\newcommand{\yonsei}{Yonsei University, IPAP, Seoul 120-749, Korea}
\affiliation{\abilene}
\affiliation{\augie}
\affiliation{\banaras}
\affiliation{\barc}
\affiliation{\bnlcoll}
\affiliation{\bnlphys}
\affiliation{\caucr}
\affiliation{\charlesczech}
\affiliation{\chonbuk}
\affiliation{\ciae}
\affiliation{\cns}
\affiliation{\colorado}
\affiliation{\columbia}
\affiliation{\czechtech}
\affiliation{\dapnia}
\affiliation{\debrecen}
\affiliation{\elte}
\affiliation{\ewha}
\affiliation{\fit}
\affiliation{\fsu}
\affiliation{\gsu}
\affiliation{\hiroshima}
\affiliation{\ihepprot}
\affiliation{\illuiuc}
\affiliation{\inrras}
\affiliation{\instpasczech}
\affiliation{\isu}
\affiliation{\jaea}
\affiliation{\jinrdubna}
\affiliation{\jyvaskyla}
\affiliation{\kek}
\affiliation{\korea}
\affiliation{\kurchatov}
\affiliation{\kyoto}
\affiliation{\labllr}
\affiliation{\lahorelums}
\affiliation{\lawllnl}
\affiliation{\losalamos}
\affiliation{\lpc}
\affiliation{\lund}
\affiliation{\maryland}
\affiliation{\mass}
\affiliation{\michigan}
\affiliation{\muenster}
\affiliation{\muhlenberg}
\affiliation{\myongji}
\affiliation{\nagasaki}
\affiliation{\newmex}
\affiliation{\nmsu}
\affiliation{\ohio}
\affiliation{\ornl}
\affiliation{\orsay}
\affiliation{\peking}
\affiliation{\pnpi}
\affiliation{\riken}
\affiliation{\rikjrbrc}
\affiliation{\rikkyo}
\affiliation{\saispbstu}
\affiliation{\saopaulo}
\affiliation{\seoulnat}
\affiliation{\stonybrkc}
\affiliation{\stonycrkp}
\affiliation{\tenn}
\affiliation{\titech}
\affiliation{\tsukuba}
\affiliation{\vandy}
\affiliation{\waseda}
\affiliation{\weizmann}
\affiliation{\wigner}
\affiliation{\yonsei}
\author{A.~Adare} \affiliation{\colorado}
\author{S.~Afanasiev} \affiliation{\jinrdubna}
\author{C.~Aidala} \affiliation{\mass} \affiliation{\michigan}
\author{N.N.~Ajitanand} \affiliation{\stonybrkc}
\author{Y.~Akiba} \affiliation{\riken} \affiliation{\rikjrbrc}
\author{H.~Al-Bataineh} \affiliation{\nmsu}
\author{J.~Alexander} \affiliation{\stonybrkc}
\author{K.~Aoki} \affiliation{\kyoto} \affiliation{\riken}
\author{Y.~Aramaki} \affiliation{\cns}
\author{E.T.~Atomssa} \affiliation{\labllr}
% ...
\collaboration{PHENIX Collaboration} \noaffiliation

\date{\today}

\begin{abstract}

The PHENIX experiment has measured the production of $\pi^0$s in Au$+$Au 
collisions at $\sqrt{s_{_{NN}}}$ = 200\,GeV. The new data offer a fourfold 
increase in recorded luminosity, providing higher precision and a larger 
reach in transverse momentum, $p_T$, to 20~GeV/$c$. The production ratio of 
$\eta/\pi^0$ is 0.46$\pm$0.01(stat)$\pm$0.05(syst), constant with $p_T$ and 
collision centrality. The observed ratio is consistent with earlier 
measurements, as well as with the $p$$+$$p$ and $d+$Au values. $\pi^0$ are 
suppressed by a factor of 5, as in earlier findings. However, with the 
improved statistical precision a small but significant rise of the nuclear 
modification factor $R_{\rm AA}$ vs $p_T$, with a slope of 
0.0106$\pm^{0.0034}_{0.0029}$~[Gev/$c$]$^{-1}$, is discernible in central 
collisions. A phenomenological extraction of the average fractional parton 
energy loss shows a decrease with increasing $p_T$. To study the path length 
dependence of suppression, the $\pi^0$ yield was measured at different 
angles with respect to the event plane; a strong azimuthal dependence of the 
$\pi^0$ $R_{\rm AA}$ is observed. The data are compared to theoretical 
models of parton energy loss as a function of the path length, $L$, in the 
medium.  Models based on pQCD are insufficient to describe the data, while a 
hybrid model utilizing pQCD for the hard interactions and AdS/CFT for the 
soft interactions is consistent with the data.

\end{abstract}

	


\maketitle


%
%
\section{Introduction}

Discovery of the suppression of high transverse momentum (\pt) hadrons
in relativistic heavy ion collisions~\cite{ppg003,ppg014,star2003} and the
absence of such suppression in \dau collisions~\cite{ppg028} inspired
intense theoretical work during the past decade.
The phenomenon was immediately interpreted, in fact, even
predicted~\cite{bjorken1982,gyulassy1992,baier2000}, as
the energy loss of a hard scattered parton in the hot, dense
strongly-interacting quark-gluon plasma (QGP) formed in the collision.
Prompted by the large amount
of very diverse experimental data from the Relativistic Heavy Ion
Collider (RHIC) -- namely, by suppression patterns at various collision
energies, colliding systems, and centralities --  several models have
been developed, based mostly on perturbative quantum chromodynamics
(pQCD) (see Section~\ref{subsec:models} as well as~\cite{brick}). 
The suppression patterns are quantified by the nuclear modification
factor \raa, defined for single-inclusive \piz{s} as

  \begin{equation}
  R_{\rm AA} (p_T) =  
  \frac{(1/N_{\rm AA}^{\rm evt}) {\rm d}^2N_{\rm AA}^{\pi^0}/{\rm d}p_T{\rm dy}}{\left<T_{\rm AB}\right>\times{\rm d}^2\sigma_{\rm pp}^{\pi^0}/{\rm d}p_T{\rm dy}},
  \end{equation}

\noindent
where $\sigma_{pp}^{\pi^0}$ is the production cross section of 
\piz in \pp collisions,  
$\left<T_{\rm AB}\right> = \left<N_{\rm coll}\right>/\sigma_{pp}^{\rm inel}$ 
is the nuclear overlap function averaged over the relevant range of impact 
parameters, and $\left<N_{\rm coll}\right>$ is the number of binary
nucleon-nucleon collisions computed with
$\sigma_{pp}^{\rm inel}$. Despite their different approaches, several
models~\cite{amy2001,wang2001,asw2003,aswads2010}
were able to describe the \pt and centrality dependence of \raa
within experimental uncertainties. At the same time,
those models provided very different estimates of 
medium properties such as the transport coefficient $\hat{q}$,
the average
4-momentum-transfer-squared per mean free path of the outgoing parton
within the medium.
For this reason, \raa alone does not provide
sufficient constraint for extracting medium properties such as
$\hat{q}$ from the theoretical predictions, because it averages the
varying energy losses along many different paths of the parton in
the medium.

While dihadron correlation measurements are a successful approach to
constrain \mean{L} of the parton in the medium~\cite{ppg090}, the
single particle observable \raa typically has smaller statistical errors and
higher \pt reach.
In addition, if \raa is
measured as a function of the azimuthal angle with respect to the
\rm event plane of the collision, 
the average path
length \mean{L} can be constrained~\cite{ppg054,ppg092}.
In all but the
most central ion-ion collisions, the overlap region of the nuclei is
not azimuthally isotropic. The average distance the parton traverses
before emerging and fragmenting varies as a function of the angle
with respect to the event plane. Each collision centrality \df class
selects different \mean{L} values, so the differential observable \raadf
directly probes the path length dependence of the energy loss.

The first measurements of azimuthal asymmetries of nuclear
suppression and collective flow~\cite{ppg054,ppg092,ppg110}
used \piz{s} as the
probe, which has the advantage that \piz{s} are relatively easy to identify
over a very wide \pt range in a single detector -- a crucial factor
in mitigating systematic uncertainties.  
As pointed out in~\cite{ppg092}, both collective flow and azimuthal
dependence of nuclear suppression can be formally defined at any
\pt, but they have historically and conceptually different roots.  The
notion of collective flow originates in lower \pt phenomena, and is
usually interpreted as a {\it boost} to the original \pt spectrum (of
partons or final state particles) in
the direction of the highest pressure gradient.  In contrast, \raa and
\raadf are typically used to describe high \pt behavior, and their
decrease from unity interpreted as a {\it loss} of parton momentum due
to the presence of a medium.
In this paper, results on \piz production, 
the nuclear modification factor \raa,
and its azimuthal dependence in terms of the
\rm event-plane-dependent \raadf are presented. 
The results presented here are
based upon the data collected in the 2007 RHIC run.  The data sample is
four times larger than that of~\cite{ppg080,ppg092}.
The dedicated reaction plane detector~\cite{nimrxnp}
installed in 2007 offers
improved \rm event-plane resolution.

\section{Experimental details}

\subsection{Data set}

\begin{figure}[htbp]
  \centering
  \includegraphics[width=1.0\linewidth]{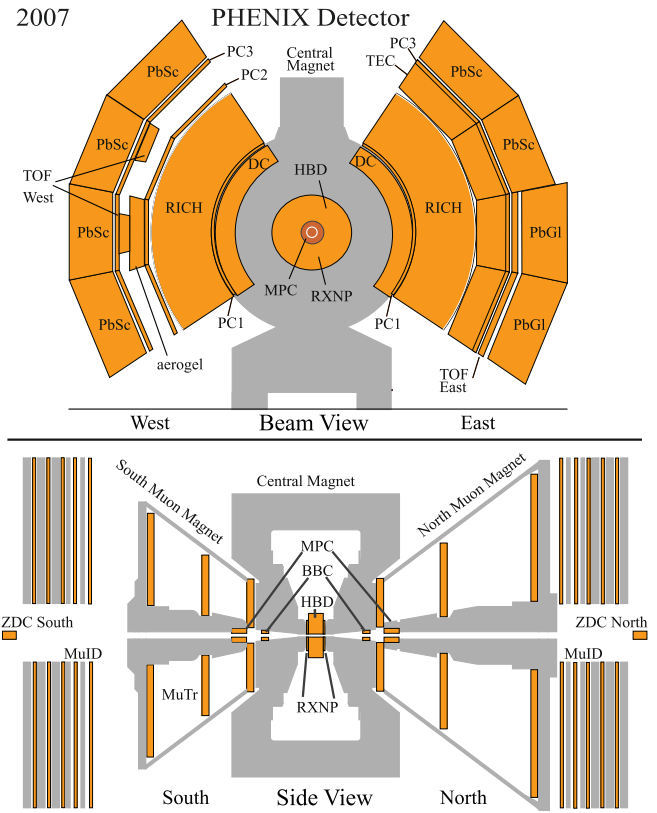}
    \caption{(Color online) 
    PHENIX experimental setup in the 2007 data taking period.
}
    \label{fig:phenix_2007}
\end{figure}

This analysis used $3.8 \times 10^9$ minimum bias \auau collisions at
$\sqrt{s_{NN}}=200$~GeV recorded by the PHENIX experiment~\cite{phenix}
at RHIC in 2007. The experimental setup is shown in
Figure~\ref{fig:phenix_2007}.
Collision centrality was determined from the amount of charge
deposited in the Beam-Beam Counters (BBC, $3.0<|\eta|<3.9$).
From a Monte Carlo calculation based on the Glauber
model~\cite{glauber1958,glauber2007}, the average
number of participants $N_{\rm part}$, the number of binary collisions
$N_{\rm coll}$, and impact parameter $b$ are estimated
(see Table~\ref{tab:glauber}).

\begingroup \squeezetable

\begin{table}[htb]
  \caption{Average \Npart, \Ncoll, impact parameter 
  and participant
    eccentricity~\cite{Alver:2006wh}  for all    centrality classes.
}
  \begin{ruledtabular}\begin{tabular}{ccccc}
    Centrality & $\langle$\Npart$\rangle$ & 
     $\langle$\Ncoll$\rangle$ &  $\langle b \rangle$ & 
     $\langle \varepsilon_{\rm part} \rangle$ \\
        (\%) &  &  & [fm] &  \\
    \hline
    00--10     &  325.8$\pm$3.8 & 960.2$\pm$96.1 & 3.1$\pm$0.1 &
    0.105 $\pm$ 0.004 \\
    10--20     &  236.1$\pm$5.5 & 609.5$\pm$59.8 & 5.6$\pm$0.2 &
    0.198 $\pm$ 0.008 \\
    20--30     &  167.6$\pm$5.8 & 377.6$\pm$36.4 & 7.3$\pm$0.3 &
    0.284 $\pm$ 0.010 \\
    30--40     &  115.5$\pm$5.8 & 223.9$\pm$23.2 & 8.7$\pm$0.3 &
    0.358 $\pm$ 0.011 \\
    40--50     &  76.2$\pm$5.5  & 124.6$\pm$14.9 & 9.9$\pm$0.4 &
    0.425 $\pm$ 0.013 \\
    50--60     &  47.1$\pm$4.7  & 63.9$\pm$9.4   & 10.9$\pm$0.4 &
    0.495 $\pm$ 0.016 \\
    60--70     &  26.7$\pm$3.7  & 29.8$\pm$5.4   & 11.9$\pm$0.5 &
    0.575 $\pm$ 0.023 \\
    70--80     &  13.7$\pm$2.5  & 12.6$\pm$2.8   & 12.6$\pm$0.8 &
    0.671 $\pm$ 0.024 \\
    80--93     &  5.6$\pm$0.8   & 4.2$\pm$0.8    & 13.9$\pm$0.5 &
    0.736 $\pm$ 0.021 \\
  \end{tabular}\end{ruledtabular}
  \label{tab:glauber}
\end{table}

\endgroup

\subsection{Reaction plane}\label{subsec:RP}
Each noncentral nucleus-nucleus collision has a well-defined
reaction plane, given by the beam direction and the impact
parameter vector of the actual collision. 
Although this reaction plane cannot be directly observed, an event plane
can be experimentally determined event-by-event using the method discussed
in detail in \cite{ppg043}.

\begin{figure}[htbp]
  \centering
  \includegraphics[width=1.0\linewidth]{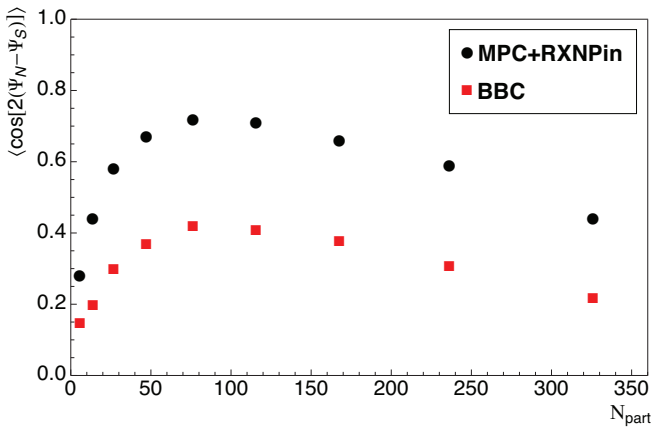}
    \caption{(Color online) Event plane resolution as a function of
  collision centrality expressed in terms of \Npart, using only
  the BBC, and using the combined MPC and RxNPin detectors. }
    \label{fig:rpres}
\end{figure}

In order to reduce the biases to the event plane determination from
physical correlations such as Hanbury-Brown-Twiss (HBT), resonance decay, and
especially high-\pt jet production, it is necessary that the event plane
is determined with a large $\eta$ gap with respect to 
the high-\pt measurement~\cite{ppg126}.
Therefore, in this
analysis measurements from two detectors were combined, located along
the beam direction to
the North and South of the interaction region.  The first is a
pair of muon-piston calorimeters (MPC)~\cite{mpc1,mpc2} covering
3.1 $<|\eta|<$ 3.9 in pseudorapidity 
and consisting of 240 2.2$\times$2.2$\times$18\,cm$^3$
$\rm{PbWO_4}$ crystals each.
The second is a pair of reaction-plane detectors (RxNP)~\cite{nimrxnp},
which are plastic scintillators, with 20\,mm of lead converter in front
of it.  The RxNP is divided into 12 azimuthal segments and further divided
radially into outer (RxNPout) and inner (RxNPin) rings.
The outer ring covers 1.0 $<|\eta|<$ 1.5 and the inner ring
covers 1.5 $<|\eta|<$ 2.8.
The current analysis did not use RxNPout and the event
plane was established only from the MPC and RxNPin.
The resolution is shown in Fig.~\ref{fig:rpres}.
The method to establish the \rm event plane from the combined
MPC-RxNPin information is identical to that used in~\cite{ppg110}.

In order to estimate the resolution of the event plane, it is measured
independently by the north and south detectors, $\Psi_N$ and $\Psi_S$,
respectively. The resolution is then characterized by
$\langle {\rm cos}[2(\Psi_N-\Psi_S)]\rangle$. 
Higher values indicate better resolution.
The resolution is centrality-dependent, as shown in \fig{fig:rpres}.

\subsection{Neutral pions}

\begin{figure}[htbp]
  \centering
  \includegraphics[width=1.0\linewidth]{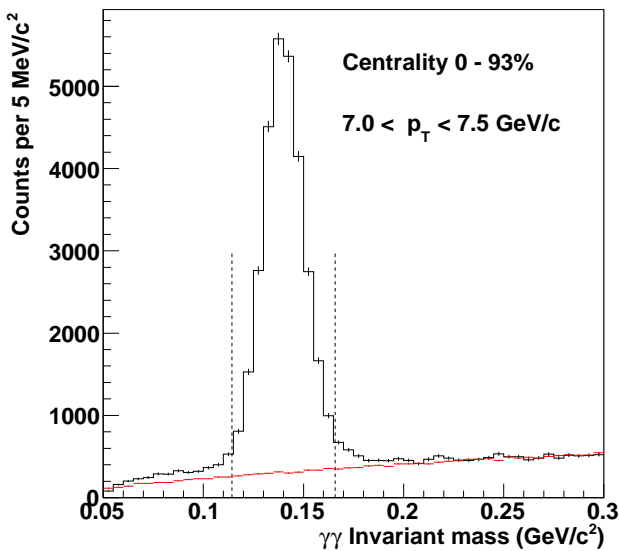}
     \caption{(Color online)
     Invariant mass spectrum of two photons (black) and
     the corresponding mixed events (red) at $7<\pt<7.5$\,\gevc in
     minimum bias collisions. Vertical lines indicate a $\pm$2.5 $\sigma$
 integration window.
}
    \label{fig:invmass}
\end{figure}

Neutral pions are measured via the $\pi^0\rightarrow\gamma\gamma$
decay channel.  Photons are identified in the PHENIX Electromagnetic
Calorimeter (EMCal, described in~\cite{nimemc}) 
consisting of two subdetectors,
both extending to $|\eta|<0.35$ in pseudorapidity and are 
located at 5.1\,m
radial distance from the collision point. The analysis uses data from the
lead-scintillator (PbSc) sampling calorimeter, which comprises six
sectors covering $3/8$ of the full azimuth and has a 5.5$\times$5.5\,cm$^2$
granularity and depth of 18 radiation lengths.
Photons are identified using various cuts on the shower shape observed
in the calorimeter as well as by comparing the observed shapes to an
ideal one, parametrized using well-controlled
test beam data~\cite{nimemc}.
Since this analysis is restricted to the \pt region above
5\,\gevc, the hadron contamination is small; hadrons in this
energy region typically deposit only a small fraction of their
energy in the EMCal. 

The invariant mass \mgg is
calculated in bins of photon pair \pt from each pair of photons,
provided the pair passes the energy asymmetry cut
$\alpha<0.8$ where 
$\alpha = |E_{\gamma_1}-E_{\gamma_2}|/(E_{\gamma_1}+E_{\gamma_2})$,
and the distance between the impact positions of the two
photons is larger than 8\,cm. An example \mgg distribution is shown in
Fig.~\ref{fig:invmass}.
For the event-plane-dependent studies the procedure is repeated in six
15$^\circ$-wide bins of angles \df with respect to the event plane.  
The combinatorial background is
estimated with the event mixing technique where photons from one event
are combined with photons from other events, which
satisfy the same global conditions (vertex position, centrality,
event plane direction), and \mgg is calculated.
The mixed-event \mgg distributions are then
normalized and subtracted from the real event distributions.
The resulting \piz peaks are $\sigma=$10-11\,MeV wide,
depending on centrality,
and have very small residual background due to the inherent
correlations in real events not reproducible by 
the mixed-event technique.  This residual background is 
fitted to a second-order polynomial in the regions below and
above the \piz peak. This polynomial shape is then subtracted from
the \mgg distribution.  
The raw \piz yields are extracted by integrating the 
resulting histogram
in a $\pm$2.5$\sigma$ wide \mgg window.

In order to establish the combined effects of acceptance and \piz
detection efficiency, 
single \piz{s} are generated with a
distribution uniform in $\phi$ and extending to $|\eta|<0.5$ in
pseudorapidity, then simulated in the full 
{\sc geant}3~\cite{GEANT}
framework of PHENIX.
After the {\sc geant}3 
output is tuned to reproduce the inactive detector
areas as well as the peak positions and widths observed in real data,
the simulated \piz{s} are embedded into real events.  The embedded
output can then be analyzed with the very same tools as the real events.

At high \pt, the two decay photons may be so close that the EMCal can
no longer resolve them as two particles and provide the proper
energies and impact points.  The two photons ``merge'' into one
cluster, and the corresponding \piz 
cannot be reconstructed from \mgg.
Such merged clusters were rejected by various shower profile cuts,
and the loss was determined by simulated \piz{s} embedded into real
events and analyzed with the same cuts.
At 11\,\gevc merging
happens only for the most symmetric decays resulting in a 5\% loss of
\piz{s}. At 17\,\gevc the correction is 50\%. At \pt= 20\,\gevc
about 70\% of \piz{s} are lost due to this effect.
The systematic uncertainties were estimated by comparing \piz yields
extracted in bins of asymmetry ($\alpha$).
The \piz yields are corrected for the \pt bin width by fitting the invariant
yield to a power-law fit and adjusting the yield to correspond to the one
at the center of the \pt bin.

\subsection{Systematic uncertainties}

\begin{table}[htb]
  \caption{Typical (minimum bias) values of systematic uncertainties 
  of the invariant yields of \piz.}
  \begin{ruledtabular}\begin{tabular}{ccccccc}
    ${\rm p_{T}}$[\gevc] & indep. & 6 & 8  & 10 &   16 & Type \\
    \hline
    Yield extr. (\%)     &  & 5.0 & 4.0 & 3.0 &  2.0 & B \\
    E scale (\%)         &  & 6.0 & 6.0 & 7.0 &  7.0 & B \\
    PID (\%)       &  & 4.0 & 3.0 & 4.0 &  5.0 & B \\
    Merging (\%)    &  &   &   & 4.5 &  28.0 & B \\
    Acceptance (\%)           &  1.0 &   &  &  &  & B \\
    Off-vertex (\%) &  1.5 &   &  &  &  & C \\
    \hline
    Total (\%)                &  1.8 & 8.8 & 7.8 & 9.7 & 29.4 & \\
  \end{tabular}\end{ruledtabular}
  \label{tab:syserr_invpi0}
\end{table}

Systematic uncertainties are characterized as follows.  Type A
uncertainties 
are point-to-point uncorrelated with \pt.
Type B uncertainties 
have point-to-point correlations that cannot be characterized
by a simple multiplicative factor, but vary smoothly with \pt.
Finally, type C uncertainties would move all points 
up or down by
a common multiplicative factor,
a typical example being the
uncertainty on \Ncoll in \raa.

The type B systematic uncertainty of the \piz raw yield extraction
has been estimated by comparing yields obtained in windows of varying widths.
The uncertainty is less
than 5\% for peripheral collisions (low multiplicity, small
combinatorics) and reaches about 7\% in central collisions.

The uncertainty on the efficiency of the photon identification 
(PID) is estimated comparing fully corrected \piz yields
obtained with various PID cuts. The uncertainty is 2--4\% at 
5--8\,\gevc, and increases both with centrality and with \pt.
It is of type B.

The uncertainty on the energy scale is estimated from how well
the peaks and widths of simulated \piz{s} embedded in real events agree
with the measured peaks and widths at each centrality.  The difference
is less than 1\% at 5--8\,\gevc.
Due to the steeply falling \piz spectrum this less-than-1\% uncertainty of
the energy scale translates to about 7\% uncertainty on the \piz 
invariant yield. 

The uncertainty due to the photon-merging correction is estimated as
follows.  Raw yields at high \pt are extracted in different asymmetry
windows both from real data and simulated decay photon pairs embedded
in real data.  Apart from small and precisely calculable acceptance
effects, the true asymmetry distribution is flat, and at any given \pt one
should observe the same raw \piz yield, for instance, in the window
$0.4<\alpha<0.6$
and $0.6<\alpha<0.8$.
However, lower asymmetry means a smaller opening angle of the decay
leading to a greater probability for the photons to merge.
Therefore, the measured asymmetry distribution at high \pt is not flat.
To determine the photon-merging correction and its systematic
uncertainty, a series of raw 
yield ratios in different asymmetry bins
were compared between data and simulation. The uncertainties on the
\piz spectra due to the merging correction are \pt and centrality
dependent.

The uncertainty due to acceptance corrections is estimated from the
ratio of simulated acceptance distribution and its fit function, which
is actually used for corrections. Since the geometry is well understood
and a single
map to exclude malfunctioning areas of the detector
has been used for the entire data set, this uncertainty 
is less than 1\% for all centralities.

There are two sources of \piz{s} not coming from the vertex (off-vertex
\piz): those produced by hadrons interacting with detector material
(instrumental background) and feed-down products from weak decay of
higher mass hadrons (physics background). Based upon simulations, both
types of background were found to be negligible  at less than 1\,\% for
\pt greater than 2.0\,\gevc,
with the exception of \piz{s} from $K^{0}_{s}$ decay
which
contribute about 3\% to the \piz yield for \pt greater than 1\,\gevc,
and have been subtracted from the data.
The uncertainty due to this effect is conservatively estimated as
1.5\,\% and is of type C.

\subsection{$R_{\rm AA}(\Delta\phi,p_{T})$}

Similar to the previous analysis~\cite{ppg092} 
the $R_{\rm AA}(\Delta\phi,p_T)$ measurement uses both
the inclusive $R_{\rm AA}(p_T)$ and the quantity
$v_2$, where $v_2$ is defined as the second Fourier expansion coefficient of
the single inclusive azimuthal distribution
\begin{equation}
\frac{dN}{d\Delta\phi} = \frac{N}{2\pi}\left(1+2v_2\cos(2\Delta\phi)\right)
\end{equation}
and $\Delta\phi = \Psi-\phi$. This assumes that the second Fourier coefficient
is dominant in this expansion. The azimuthal anisotropy $v_{2}$ has
been published in \cite{ppg110}. 

The \piz yield is subdivided into six evenly-spaced azimuthal bins
in $\Delta\phi$ from 0 to $\pi/2$ on an event-by-event basis using the
measured event plane (see Sec.~\ref{subsec:RP}). From the inclusive
$R_{\rm AA}$ the $\Delta\phi$-dependent $R_{\rm AA}$ can be constructed as
\begin{eqnarray}
R_{\rm AA}(\Delta\phi_{{\rm i}},p_{T}) &=& F(\Delta\phi_{{\rm i}},p_{T})\times R_{\rm AA}(p_{T}),
\label{eq:RaaPhi_1}
\end{eqnarray}
where
\begin{eqnarray}
F(\Delta\phi_{\rm i},p_{T}) &=&
\frac{N(\Delta\phi_{\rm i},p_{T})}
{\frac{1}{n}\displaystyle
\sum^{n}_{\rm i=1}N(\Delta\phi_{\rm i},p_{T})},
\label{eq:RaaPhi_2}
\end{eqnarray}
and the summation runs over the $n=6$ azimuthal bins.

Because of finite event plane resolution, 
$F(\Delta\phi_{\rm i},p_{T})^{\rm meas}$, as
calculated from the raw yields, needs to be corrected. An approximate unfolding
can be done by using the raw $v_2^{raw}$ and the resolution-corrected
$v_2^{\rm corr}$

\begin{eqnarray}
 F(\Delta\phi_{\rm i},p_{T}) &=&  \nonumber \\
F(\Delta\phi_{\rm i},p_{T})^{\rm meas} 
 \times
\frac{1+2v_{2}^{\rm corr}\cos(2\Delta\phi)}
{1+2v_{2}^{\rm raw}\cos(2\Delta\phi)}.
\label{eq:RaaPhi_3}
\end{eqnarray}

\begin{figure}[htbp]
  \includegraphics[width=1.0\linewidth]{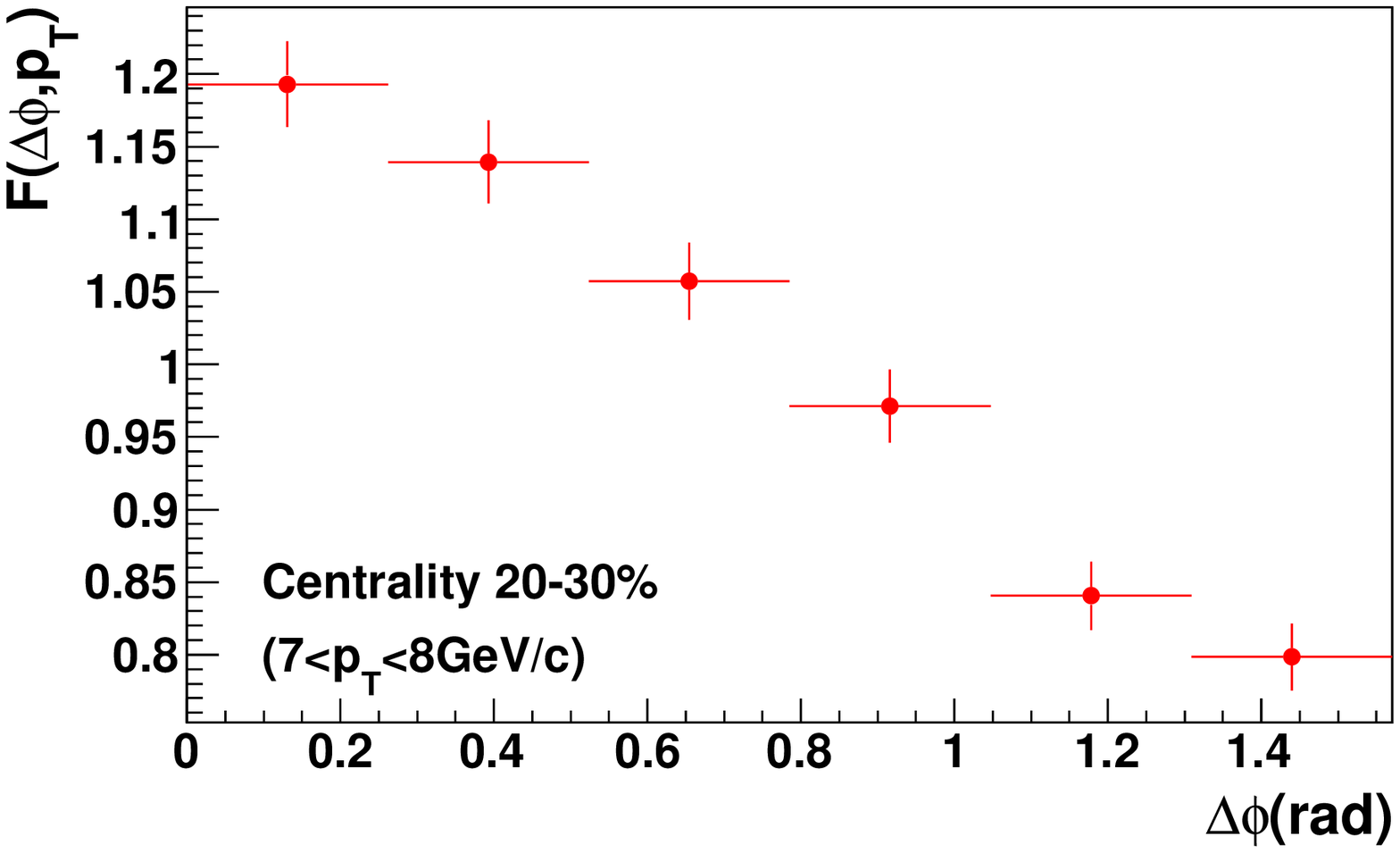}
    \caption{(Color online)
The corrected ratio, $F(\Delta\phi,p_{T})$, as a function of
azimuthal angle at centrality 20--30$\%$.
}
    \label{fig:CorrFac}
  \includegraphics[width=0.97\linewidth]{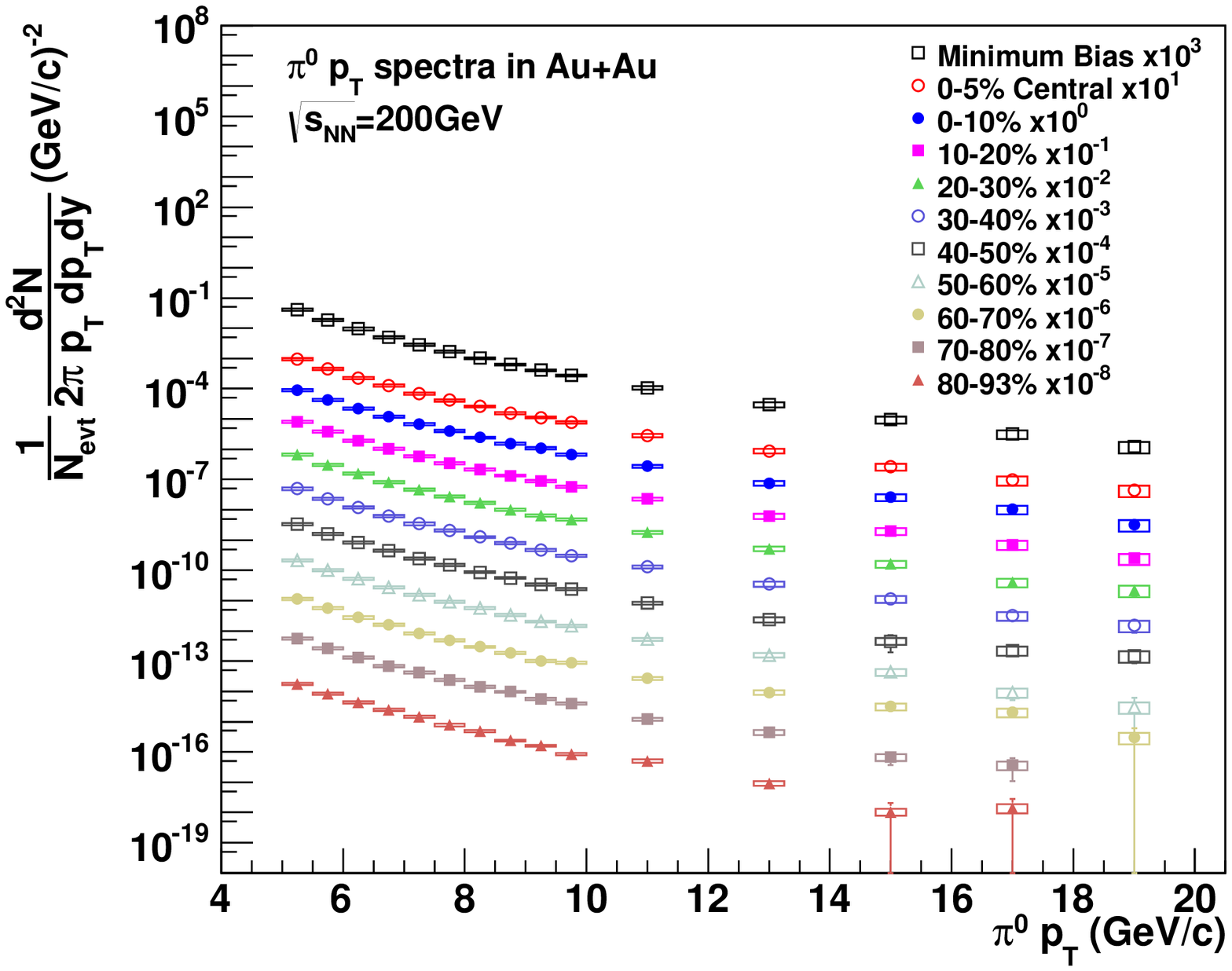}
    \caption{
    (Color online) Invariant yield of \piz as a function of \pt
    for each 10\% centrality class, 0--5\% centrality
    and minimum bias.  The \pt-scale starts at 4\,\gevc.
    Error bars are the sum of statistical and
    type A systematic uncertainties, boxes are the sum of type B and C
    systematic uncertainties.
}
    \label{fig:invyield}
\end{figure}

The relation between the raw and the corrected $v_2$ is given by
\begin{eqnarray}
v_2^{\rm cor} & = & \frac{v_2^{raw}}{\left<\cos[2(\Psi_N-\Psi_S)]\right>}.
\end{eqnarray}
The denominator is shown in Fig.~\ref{fig:rpres}.
Figure~\ref{fig:CorrFac} shows the $F(\Delta\phi,p_{T})$ at
$7<p_{T}<8$\,\gevc for centrality 20--30$\%$.

\section{Results}

\begin{figure}[htbp]
  \centering
  \includegraphics[width=1.0\linewidth]{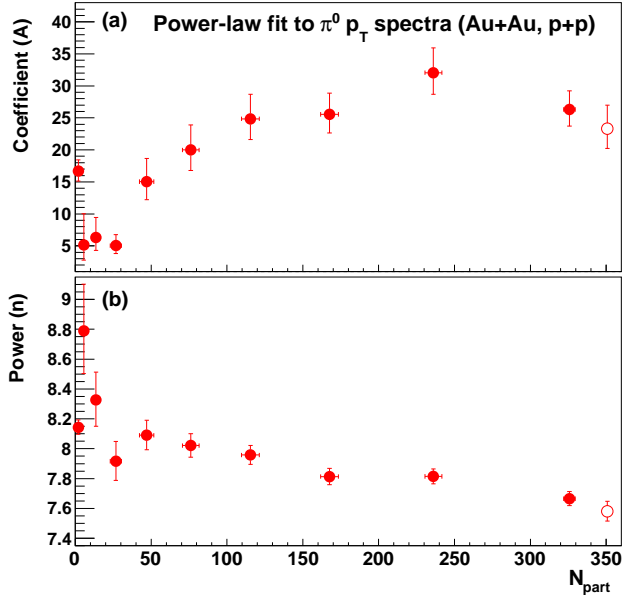}
    \caption{
    (Color online)  Power law fit parameters (as tabulated in
      Table~\ref{tab:specfitpowers7}) to the \piz spectra in the
      7--20\,\gevc \pt range as a function of centrality, expressed in
      terms of \Npart and for \pp (first point). 
      Shown are (a) amplitudes, (b) powers.
    Note that the open points (\Npart=\,352) correspond to 0--5\,\%
    centrality and partially overlap with the adjacent points
    (\Npart=\,325, 0--10\,\%).
    }
    \label{fig:fitparspectra}
\end{figure}

\subsection{Spectra and power law fits}

\begin{table}[htbp]
\caption{Fit parameters of the power law fit 
$f(p_T) = Ap_T^{-n}$ to the invariant yield
(7 $<p_T<$ 20\,\gevc range) in various centrality \auau collisions and
the \pp cross section~\cite{ppg063}.}
\begin{ruledtabular}\begin{tabular}{cccc}
System & A & n & $\chi^2$/NDF \\
\hline
\auau 0--5\,\% & 23.3$^{+3.67}_{-3.11}$ & 7.58$\pm$0.07 & 7.36/9\\
\auau 0--10\,\% & 26.3$^{+2.9}_{-2.6}$ & 7.66$\pm$0.05 & 5.43/9\\
\auau 10--20\,\% & 32.1$^{+3.9}_{-3.4}$ & 7.81$\pm$0.05 & 1.38/9\\
\auau 20--30\,\% & 25.6$^{+3.3}_{-2.9}$ & 7.81$^{+0.06}_{-0.05}$ & 14.2/9\\
\auau 30--40\,\% & 24.9$^{+3.9}_{-3.3}$ & 7.96$\pm$0.06 & 11.3/9\\
\auau 40--50\,\% & 20.0$^{+3.9}_{-3.2}$ & 8.02$\pm$0.08 & 7.50/9\\
\auau 50--60\,\% & 15.0$^{+3.6}_{-2.8}$ & 8.09$\pm$0.10 & 5.56/9\\
\auau 60--70\,\% & 5.04$^{+1.73}_{-1.24}$ & 7.92$\pm$0.13 & 12.6/9\\
\auau 70--80\,\% & 6.32$^{+3.12}_{-2.02}$ & 8.33$^{+0.19}_{-0.18}$ & 6.48/8\\
\auau 80--93\,\% & 5.16$^{+4.85}_{-2.38}$ & 8.79$^{+0.31}_{-0.29}$ & 8.14/8\\
\auau 0--93\,\% & 16.4$^{+0.93}_{-0.87}$ & 7.86$\pm$0.02 & 11.2/9\\
\pp ($\sigma$)  & 16.7$^{+1.73}_{-1.55}$ & 8.14$\pm$0.05 & 15.9/9\\
\end{tabular}\end{ruledtabular}
\label{tab:specfitpowers7}
\end{table}

\fig{fig:invyield} shows the \piz invariant yield in \auau collisions for
all centralities, and for minimum bias data.  As with earlier published \piz
results~\cite{ppg080}, in this \pt range all distributions are 
well described by a single power law 
function [$f(p_T) = Ap_T^{-n}$]. The fit method employed here
takes both statistical and systematic uncertainties into account, 
following the one established in previous
publications~\cite{ppg079,ppg080,ppg115}. The obtained fit parameters
are listed in 
Table~\ref{tab:specfitpowers7} 
for all \auau centrality classes, as well as for \pp measured in
2005~\cite{ppg063}. In the more peripheral
collisions the \auau and \pp powers are consistent, but in central
collisions the \auau powers are slightly smaller, which is
also reflected in the behavior of the nuclear modification factor (see
Sec.~\ref{sec:raa}). 
Figure~\ref{fig:fitparspectra} shows the amplitudes
and powers from Table~\ref{tab:specfitpowers7}.

\subsection{The production ratio $\eta$/\piz}

\begin{figure}[htbp]
  \centering
  \includegraphics[width=1.0\linewidth]{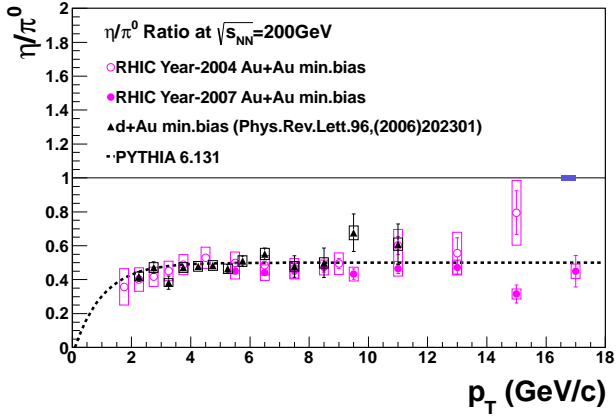}
  \caption{(Color online) 
  The $\eta$ to $\pi^{0}$ ratio as a function of $p_T{}$
  for minimum bias \auau collisions in 2004 and 2007 (this
  analysis) data sets, \dau collisions (2003~\cite{ppg051}), and
  {\sc pythia}~6.131~\cite{sjostrand2001}. 
  Boxes are \pt-correlated
  systematic uncertainties (type B), the shaded box at one is
  the global uncertainty (type C). 
}
  \label{fig:eta2pi0_all}
\end{figure}

\begin{figure}[htbp]
  \centering
  \includegraphics[width=1.0\linewidth]{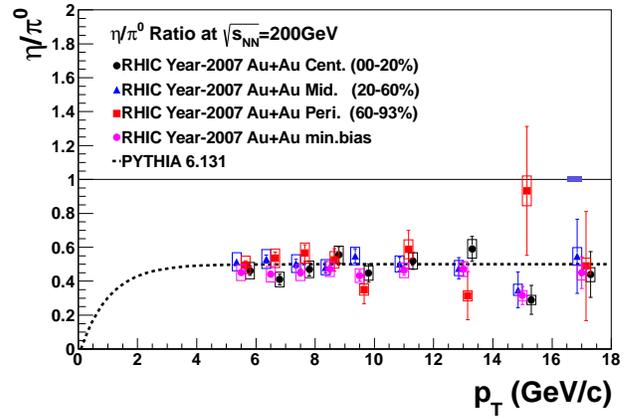}
  \caption{(Color online)
  The centrality dependence for the ratio of $\eta$ to
  $\pi^{0}$ as a function of $p_T{}$ and the expectation from
  {\sc pythia}~6.131~\cite{sjostrand2001}. 
  Boxes are \pt-correlated
  systematic uncertainties (type B), the shaded box at one is
  the global uncertainty (type C).
}
  \label{fig:eta2pi0_cent}
\end{figure}

\begin{figure}[htbp]
  \centering
  \includegraphics[width=1.0\linewidth]{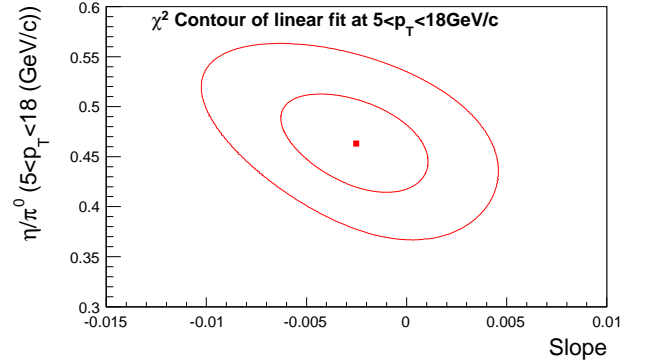}
  \caption{(Color online) One and two sigma standard deviation
  $\chi^2$ contours of the linear fit to the minimum bias $\eta$/\piz
  ratio. 
}
  \label{fig:eta2pi0_fit_contour}
\end{figure}

Combining the current high statistics \piz results with the published
$\eta$ meson spectra from the same (2007) data set~\cite{ppg115} provides
new $\eta/$\piz ratios with much smaller uncertainties than those
published previously~\cite{ppg051}. Figure~\ref{fig:eta2pi0_all} compares
the measured $\eta/$\piz ratios from minimum bias collisions 
for various data sets and colliding systems.  Although the uncertainties
vary, the new ratios are consistent with previously published
ones~\cite{ppg051} and are also consistent with the overlaid
{\sc pythia}-6.131 \pp calculation.

\begin{figure}[htbp]
  \centering
  \includegraphics[width=0.75\linewidth]{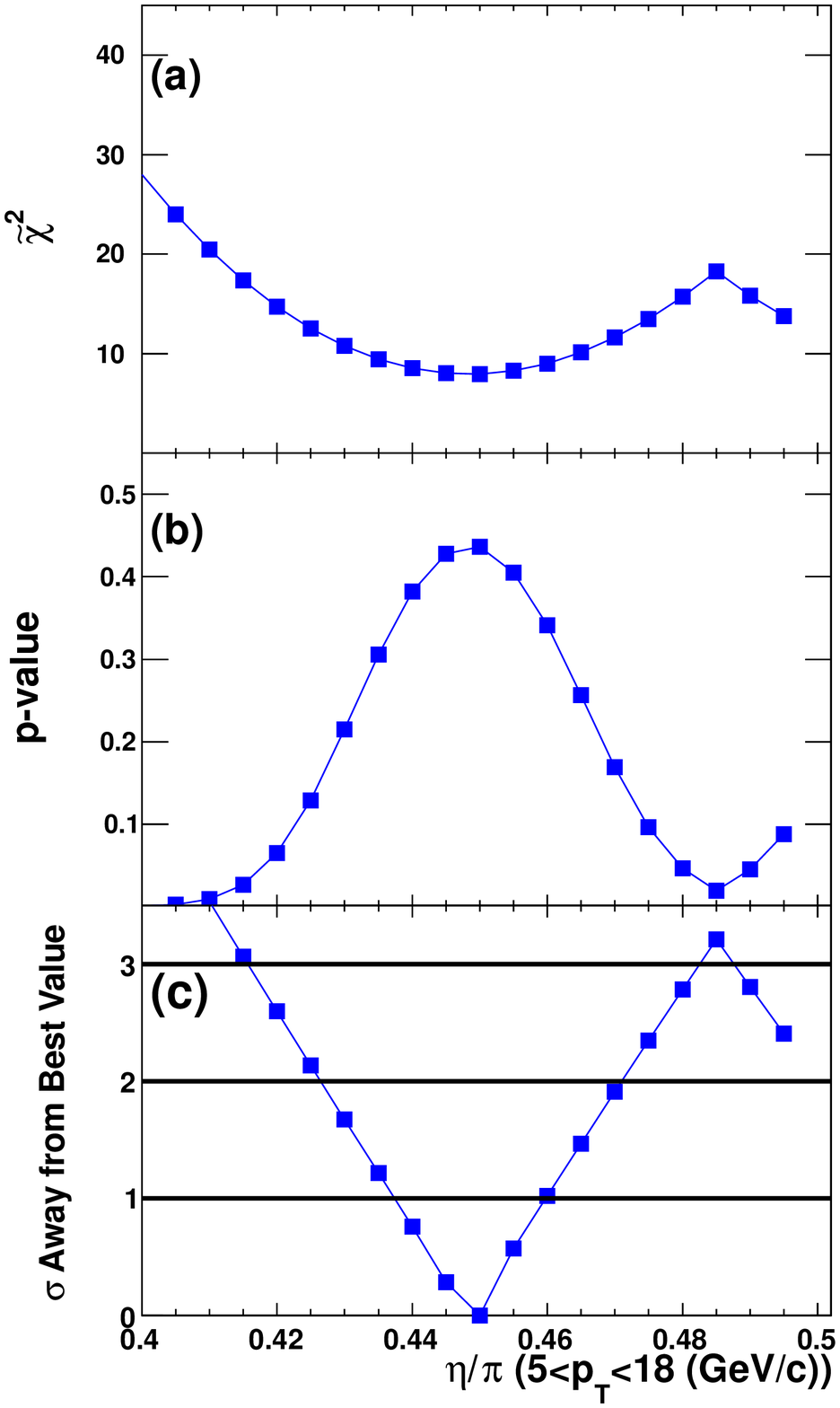}
  \caption{(Color online) Statistical analysis of the constant fit to
  the minimum bias $\eta$/\piz ratio following the method
  in~\cite{ppg079}. 
}
  \label{fig:eta2pi0_fitresult}
\end{figure}

Figure~\ref{fig:eta2pi0_cent} shows the $\eta/$\piz ratios for
various centralities along with the 
{\sc pythia} \pp values.
A linear fit to the minimum bias data gives a constant term of
0.46$\pm$0.05 and a slope of -0.0025$\pm$0.0037, with the
$\chi^2$ contours shown in Fig.~\ref{fig:eta2pi0_fit_contour}.
The fit method employed here takes both statistical and
systematic uncertainties into account, following the one
established in previous publications~\cite{ppg079,ppg080,ppg115},
and fit values for all
centralities are listed in Table~\ref{tab:etapi0fit}.
Since the data are fully consistent with a zero slope, they were
refitted with a constant in the 5--18\,\gevc \pt range, 
resulting in the final values of 
$\eta$/\piz = 0.45$^{+0.01}_{-0.01}$ for minimum bias,
$\eta$/\piz = 0.47$^{+0.01}_{-0.02}$ for 0--20\,\%,
$\eta$/\piz = 0.51$^{+0.01}_{-0.01}$ for 20--60\,\%,
and $\eta$/\piz = 0.51$^{+0.02}_{-0.02}$ for 60--93\,\% centrality.
Results of the statistical analysis of the constant fit to the minimum
bias data are shown in Fig.~\ref{fig:eta2pi0_fitresult}.
Note that the earlier
published value~\cite{ppg051} for the most central \auau collisions
was $\eta/$\piz=0.40$\pm$0.04; the current result is closer to the
$\eta/$\piz ratios observed in \dau (0.47$\pm$0.03) and \pp
(0.48$\pm$0.03)~\cite{ppg051}.

\begin{table}[htbp]
\caption{Fit parameters of linear fit to the $\eta/$\piz ratio in
  200\,\gevc \auau collisions of various centralities.
}
\begin{ruledtabular}\begin{tabular}{cccc}
Centrality & Intercept & Slope [1/\gevc] & $\chi^2$/NDF \\
\hline
00-93\,\% & 0.463$\pm$0.049 & -2.52$\times10^{-3}\pm3.66\times10^{-3}$ & 7.46/7\\
00-20\,\% & 0.463$\pm$0.053 & 3.33$\times10^{-3}\pm5.76\times10^{-3}$ & 14.8/7\\
20-60\,\% & 0.525$\pm$0.058 & -5.67$\times10^{-3}\pm5.43\times10^{-3}$ & 4.03/7\\
60-93\,\% & 0.511$\pm$0.061 & -2.80$\times10^{-3}\pm1.03\times10^{-2}$ & 9.36/7\\
\end{tabular}\end{ruledtabular}
\label{tab:etapi0fit}
\end{table}

The lack of nuclear effects on this ratio indicate that at high p$_T$
the fragmentation occurs outside the medium and the ratio is governed
by vacuum fragmentation~\cite{ppg051}.
This is also supported by a recent global
analysis of $\eta$ fragmentation functions (consider Figure 5
in~\cite{aidala2011} and the fact that the relevant $z$ range, the
fraction of the four-momentum of the parton taken by a fragment, in
the current measurement is about 0.05--0.2).
The relevant p$_T$ is presumably 5--6\,\gevc, below which
recombination may be a significant hadronization mechanism
(see~\cite{ppg123,Afanasiev:2007tv} and~\cite{bass2009}).
Also, it should be pointed out, that precise knowledge of the
absolute value of this ratio is important for the background
calculations in dielectron and direct photon measurements.

\begin{figure*}[htbp]
  \includegraphics[width=1.0\linewidth]{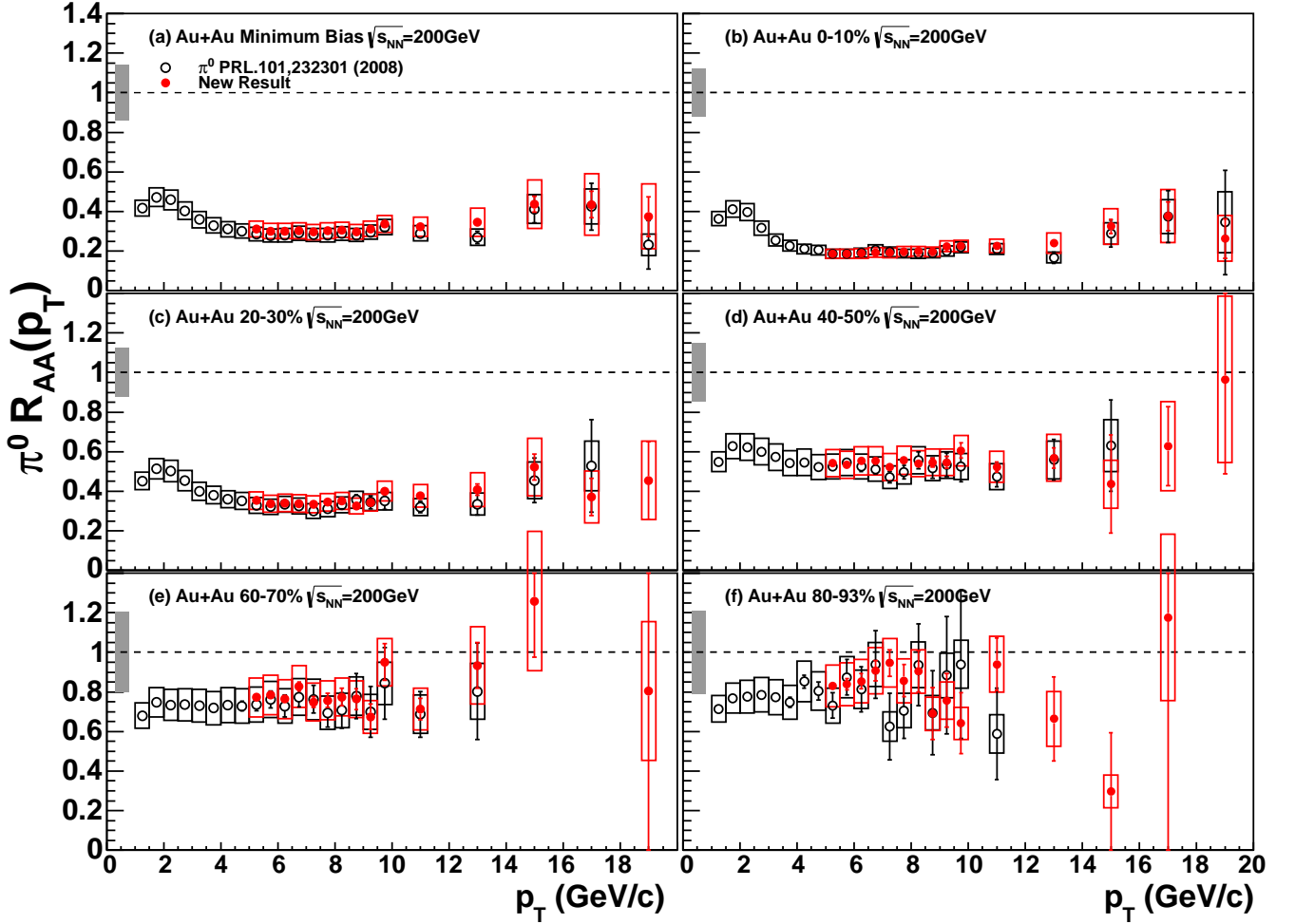}
  \caption{(Color online) 
  The nuclear modification factor \raa of \piz as a function
  of \pt for various 10\%-wide centrality classes.
  Closed (red) circles are the results from the current analysis,
  while open (black) circles are the data published in~\cite{ppg080}.
  Shaded (gray) boxes around 1 indicate global systematic 
  uncertainties and
  are of type C.  The \pp reference is from the 2005 PHENIX
  data~\cite{ppg063}.
  }
    \label{fig:run7raa}
\end{figure*}

\subsection{Nuclear modification factor ($\phi$-integrated)}
\label{sec:raa}

\begin{figure}[htbp]
  \includegraphics[width=1.0\linewidth]{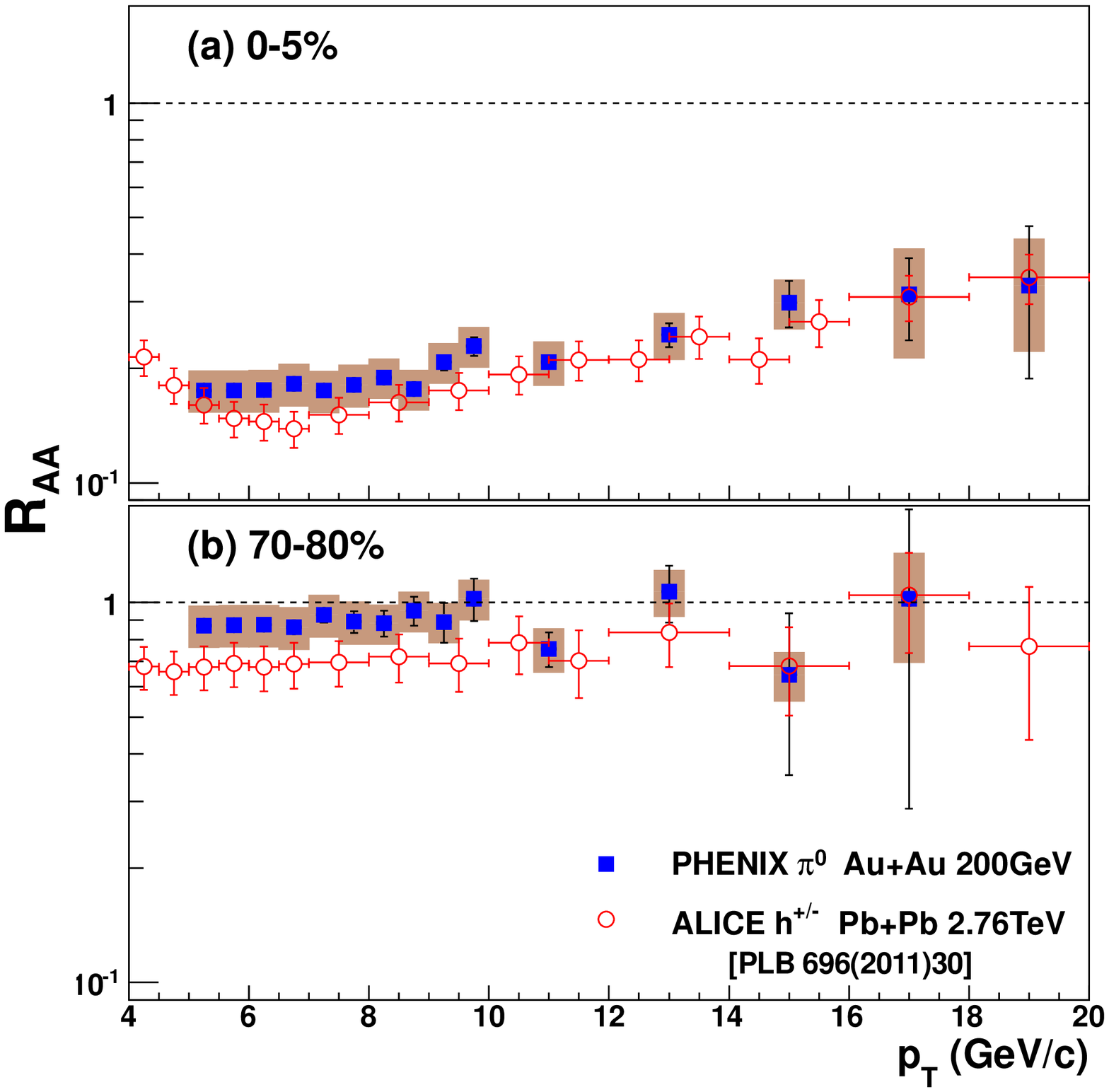}
  \caption{(Color online)
  Comparison of the \piz \raa from this measurement and the
  charged hadron \raa in \pbpb collisions at \sqsn= 2.76\,TeV 
  from the ALICE experiment~\cite{alice2011} at LHC. 
  The central and peripheral classes are (a) 0-5\% and (b) 70-80\%.
  }  
  \label{fig:phenixalicecomp}
\end{figure}

The reference yield of $\pi^{0}$ in \pp collisions has been
obtained from data taken in 2005~\cite{ppg063}. Instead of using a fit
to the \pp data, \raa has been calculated by dividing the
\auau yields point-by-point by the \TAB-scaled \pp cross section.
Figure~\ref{fig:run7raa} shows \raa for \piz{s} as a function of 
$p_{T}$ for six representative centrality classes with the new results
overlaid on the previously published ones~\cite{ppg080}.  
The analysis presented here spans the range \pt = 5--20\,\gevc
in several centrality classes.
Gray bands show the global systematic uncertainties and are of
type C, which are the quadratic sum of uncertainties of \Ncoll,
\pp normalization, and off-vertex \piz contribution shown in
Table~\ref{tab:syserr_invpi0}.
The results agree well in the overlapping \pt region with the 
published \raa data~\cite{ppg080}.

Figure~\ref{fig:phenixalicecomp} compares 
current RHIC \sqsn= 200\,GeV \auau 
\piz \raa data to the charged hadron \raa observed in
\sqsn= 2.76\,TeV \pbpb collisions at the LHC (ALICE
experiment)~\cite{alice2011}.  For the \pbpb points, the vertical
error bars show the total errors.
For both centralities and over the entire \pt range of 5--20\,\gevc,
the two data sets appear to be
similar. This is remarkable given the
14-fold increase of colliding energy,
resulting in an approximately factor of
two increase in the parton density at the LHC~\cite{horowitz2011}.
The expected increase in the parton density is corroborated by the factor
of 2.2 increase in dN$_{ch}$/d$\eta$ reported by ALICE~\cite{Aamodt:2010pb}.

However, there are two important caveats.  
Preliminary results from the same experiment on \piz{s}, measured 
via photon conversions up to 10\,\gevc, show an
\raa that is somewhat lower in central collisions than for charged
hadrons~\cite{appelshauser2011}.  
In~\cite{horowitz2011} the authors assert that the similarity of 
\raa at RHIC and LHC may be coincidental.
In any case, it does not mean that the RHIC and LHC data show the
same average parton energy loss $\langle \varepsilon \rangle$ (see
Sec.~\ref{sec:eloss}), since the spectra are much harder
(the power $n=6$) at the LHC. The power is obtained by fitting
the ALICE charged hadron data~\cite{alice2011}.

The fact that at \sqsn= 200\,GeV in central collisions \raa reaches 
its minimum around 5\,\gevc transverse momentum 
was first observed in~\cite{ppg014}.
higher \pt \raa appeared to be approximately constant, although the data
did not unambiguously exclude a slow rise with 
\pt~\cite{ppg079,ppg115}.  On the other hand, all models that reproduce
the large suppression observed at \pt of 6--10\,\gevc predict a slow rise
of \raa as the transverse momentum increases~\cite{majumder2011,horowitz2011}.
The current, higher precision data are used to reassess the
\pt-dependence of \piz suppression in the RHIC regime.  

\begin{figure}[htbp]
  \centering
  \includegraphics[width=1.0\linewidth]{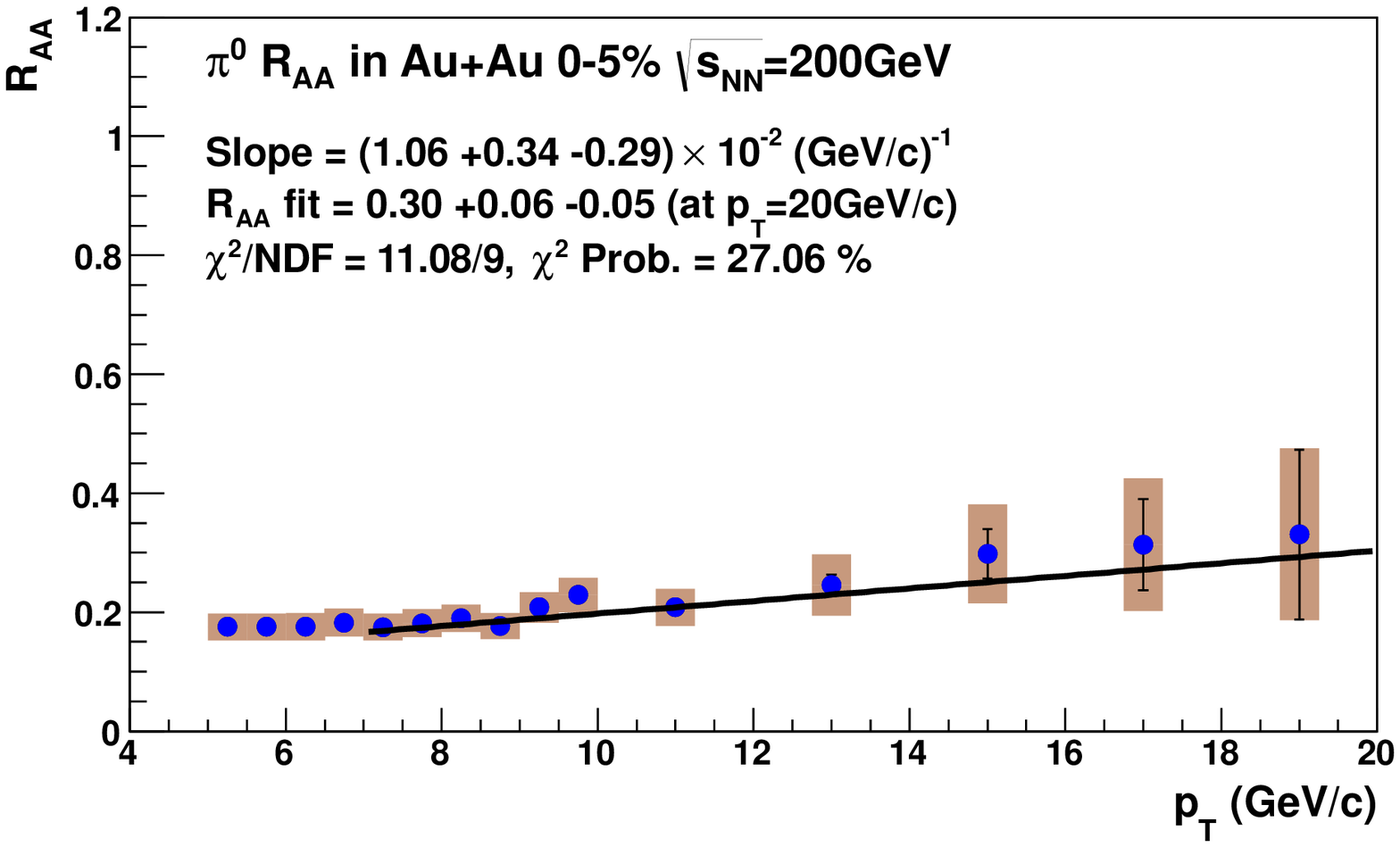}
  \caption{(Color online) 
  Linear fit to the \pt-dependence of \raa in the 7--20\,\gevc
  \pt range  in the most central (0--5\%) \auau collisions.  Both
  statistical (bars) and systematic (boxes) uncertainties are considered
  in the fit.
  }
    \label{fig:RaaFitCentDep}
\end{figure}

\begin{figure}[htbp]
  \centering
  \includegraphics[width=1.0\linewidth]{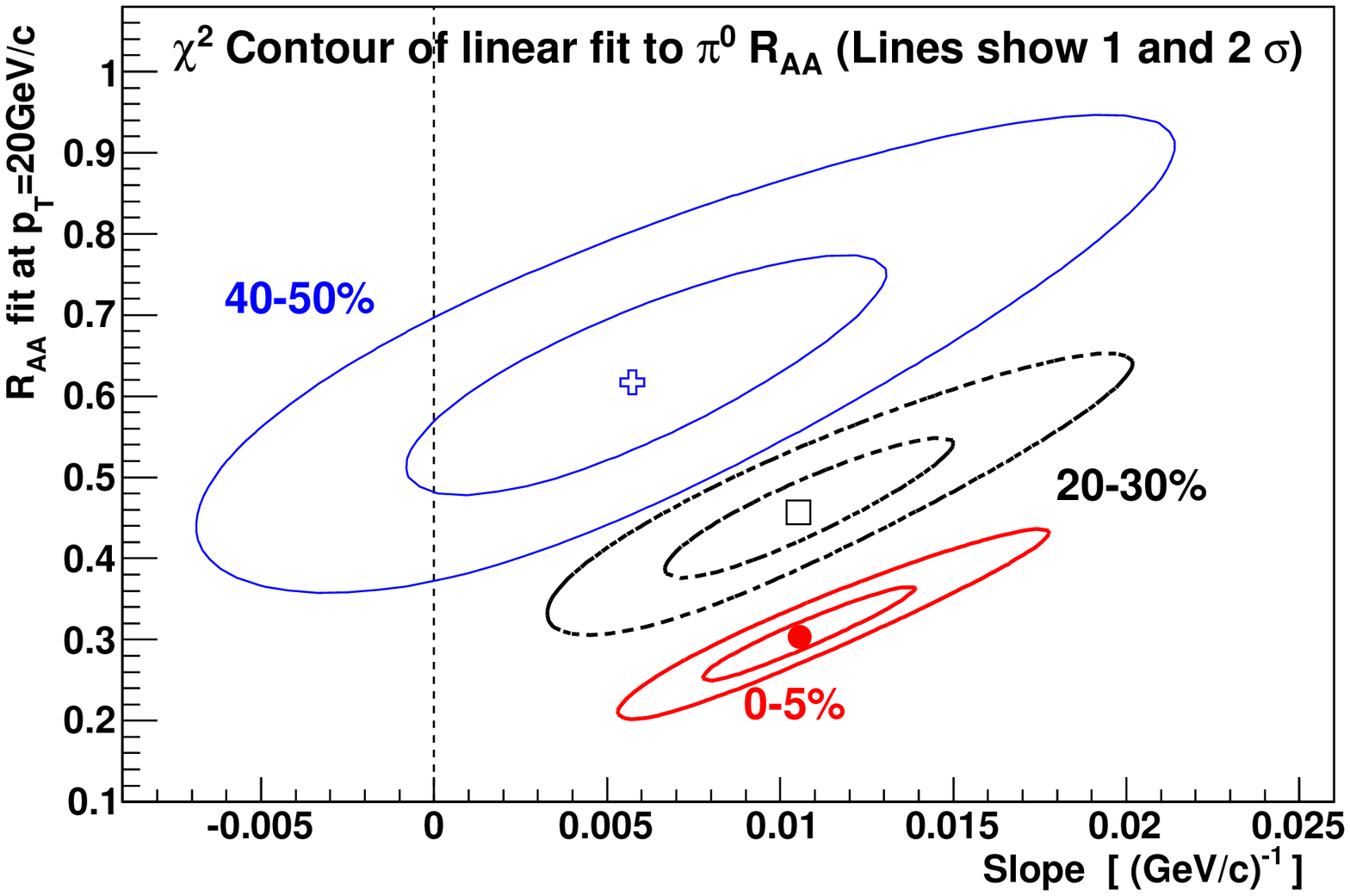}
  \caption{(Color online) 
  One and two sigma standard deviation $\chi^2$ contours of the linear fit 
  (cf. Figure~\ref{fig:RaaFitCentDep}) to
  the \pt-dependence of \raa for three different centralities.
  }
    \label{fig:ContourTwoSigPi0}
\end{figure}

Figure~\ref{fig:RaaFitCentDep} shows a sample linear fit to the 
\pt-dependence
of \raa in the most central \auau data.  
Figure~\ref{fig:ContourTwoSigPi0} shows the 1 and
2$\sigma$ contour lines of the fitted slope and intercept for three
centralities.
The fit method employed here takes both statistical and
systematic uncertainties into account, following the one
established in previous publications~\cite{ppg079,ppg080,ppg115}.
In contrast to Figure 9 in~\cite{ppg079} where
the slope was consistent with zero within 1$\sigma$ due to the large
uncertainties, the slope here is significantly different from zero, not 
only in the most central, but in 20--30\% centrality collisions as well.

\begin{figure}[htbp]
  \centering
  \includegraphics[width=1.0\linewidth]{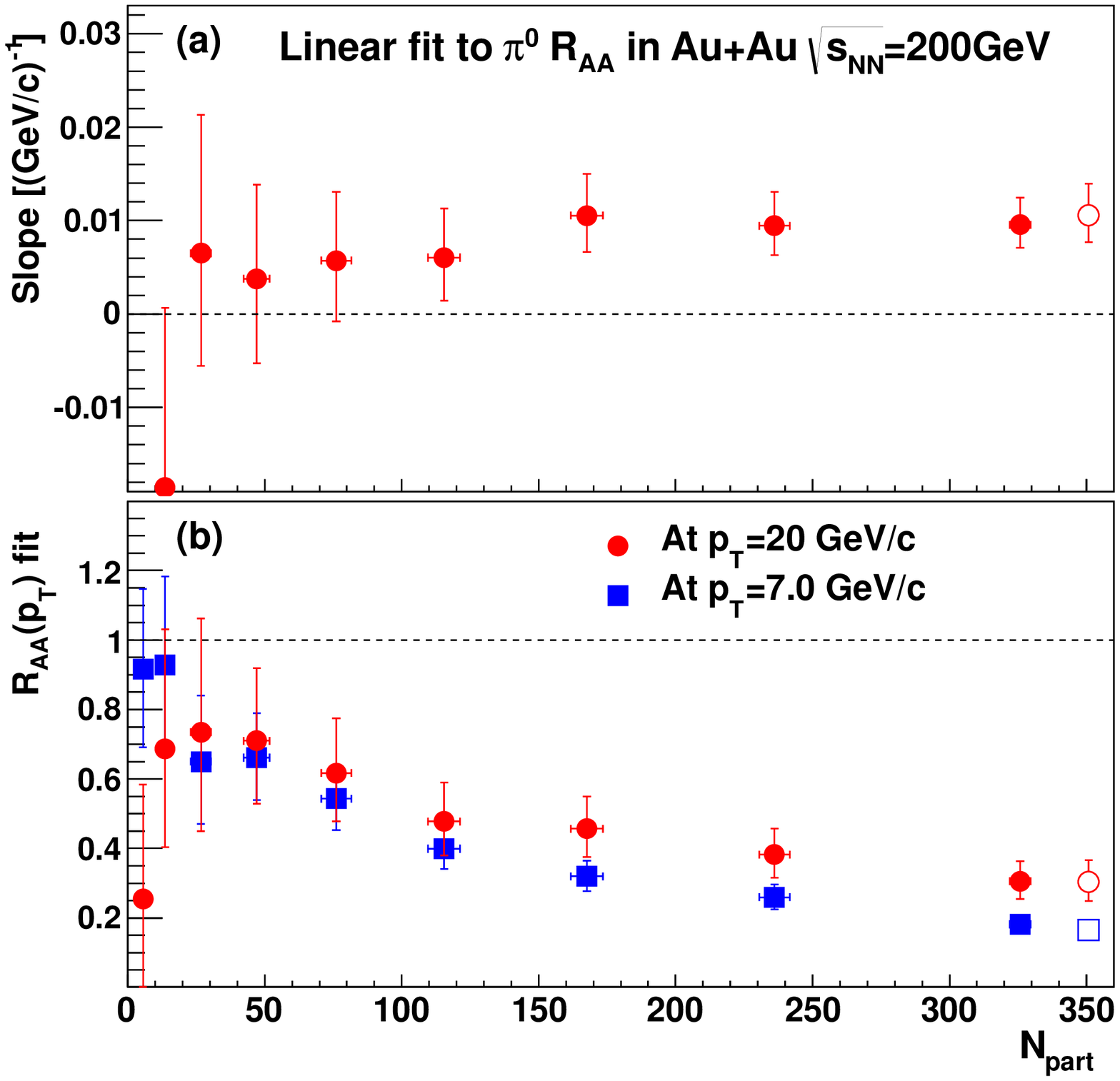}
  \caption{(Color online)
  (a) Slopes of the linear fits to \piz \raa vs \pt 
  in \sqsn=200\,GeV \auau collisions
  (as shown in Fig.~\ref{fig:RaaFitCentDep}) and
  the fitting uncertainties.  The fits are in the 7--20\,\gevc \pt
  range. The horizontal axis is centrality, expressed in
  terms of \Npart.
  (b) Values of \raa calculated from the fits at \pt= 7\,\gevc and
  \pt= 20\,\gevc, also as a function of centrality.  
  Note that the open points (\Npart= 352) correspond to 0--5\,\%
  centrality and partially overlap with the adjacent points
  (\Npart= 325, 0--10\,\%).
  }
    \label{fig:SlopeOffsetPi0Raa}
\end{figure}

Figure~\ref{fig:SlopeOffsetPi0Raa} shows the fitted slopes (a) 
and the \raa from the fits (b) at 7\,\gevc and 20\,\gevc for all
centralities, expressed in terms of \Npart.
At and above \Npart=167 (20--30\% centrality) the slopes are significantly
different from zero.

\subsection{Phenomenological energy loss}
\label{sec:eloss}

\begin{figure}[htbp]
  \centering
  \includegraphics[width=1.0\linewidth]{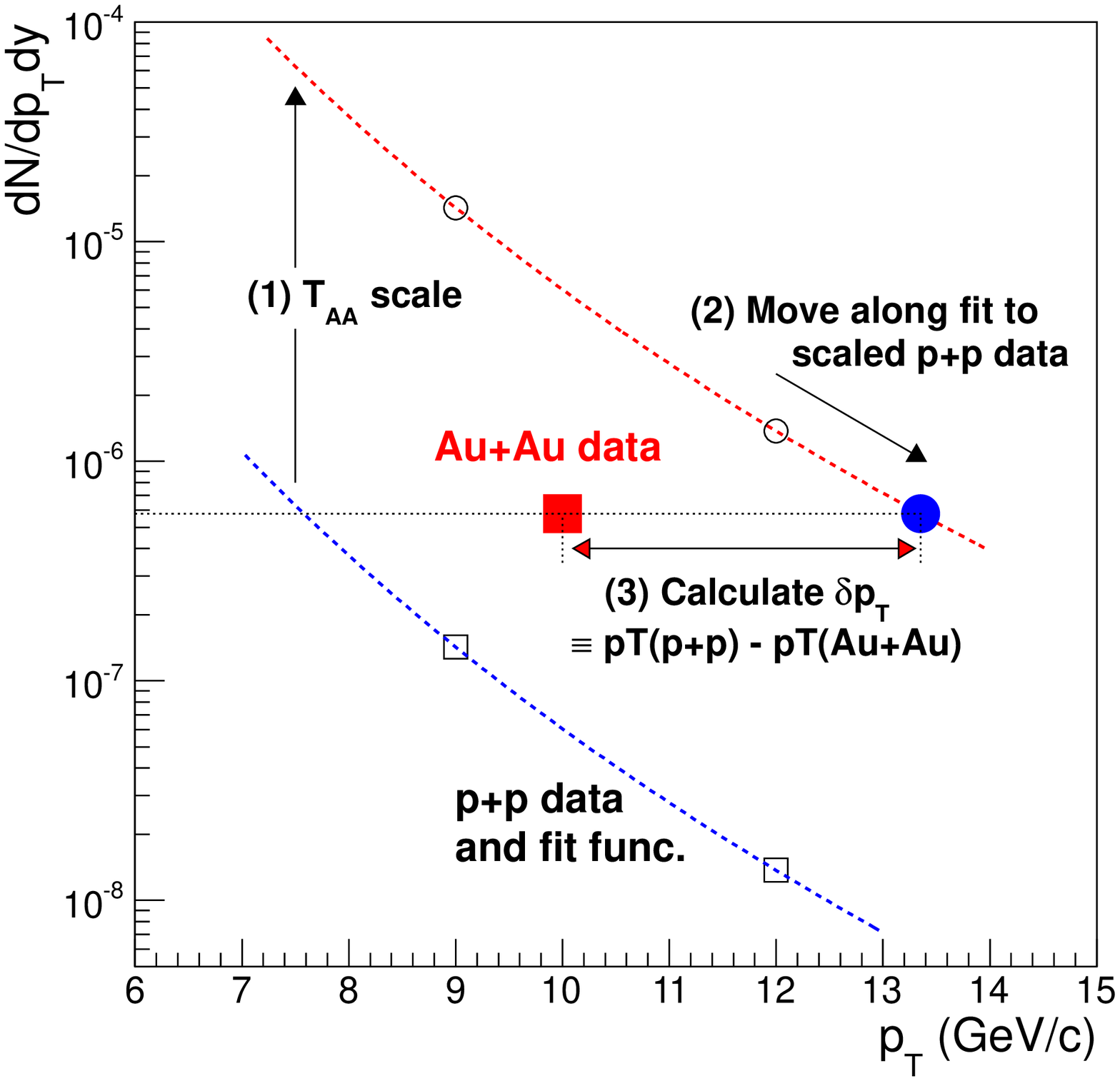}
  \caption{(Color online) 
  Method of calculating average fractional momentum loss
  ($S_{{\rm loss}} \equiv \delta p_T/p_T$).
  Figure is for illustration only, and errors are not shown.
  In the order of procedure: (1) Scale the \pp data by \taa corresponding
  to centrality selection of \auau data, (2) shift the \pp points closest
  to \auau in yield, and (3) calculate momentum difference of \pp and
  \auau points.
  }  
  \label{fig:slossmethod}
\end{figure}

\begin{figure}[htbp]
  \centering
  \includegraphics[width=1.0\linewidth]{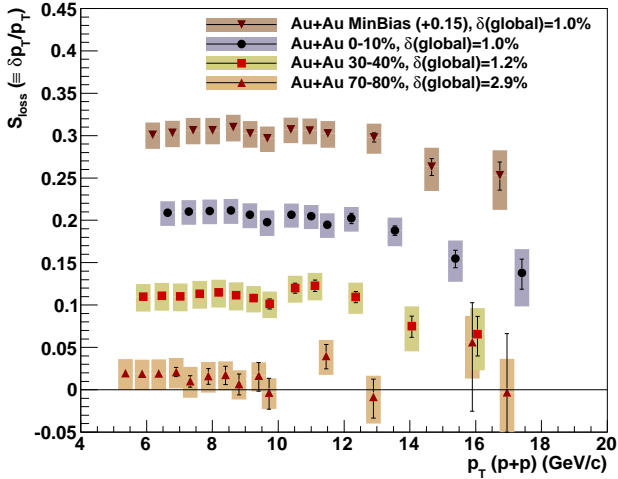}
  \caption{(Color online) 
  Average fractional 
  momentum loss, as defined in the text, between various
  centrality \auau and \taa-scaled \pp collisions.  The horizontal
  axis is the \pt in the \pp collision.  Note that for clarity the
  minimum bias data are shifted up by 0.15.
  $\delta$(global) stands for the uncertainty coming from the
   uncertainties of \taa. The overall normalization error from the
   \pp measurement is 1.3\,\%, and is not shown here.
  }  
  \label{fig:sloss}
\end{figure}

\begin{figure}[htbp]
  \centering
  \includegraphics[width=1.0\linewidth]{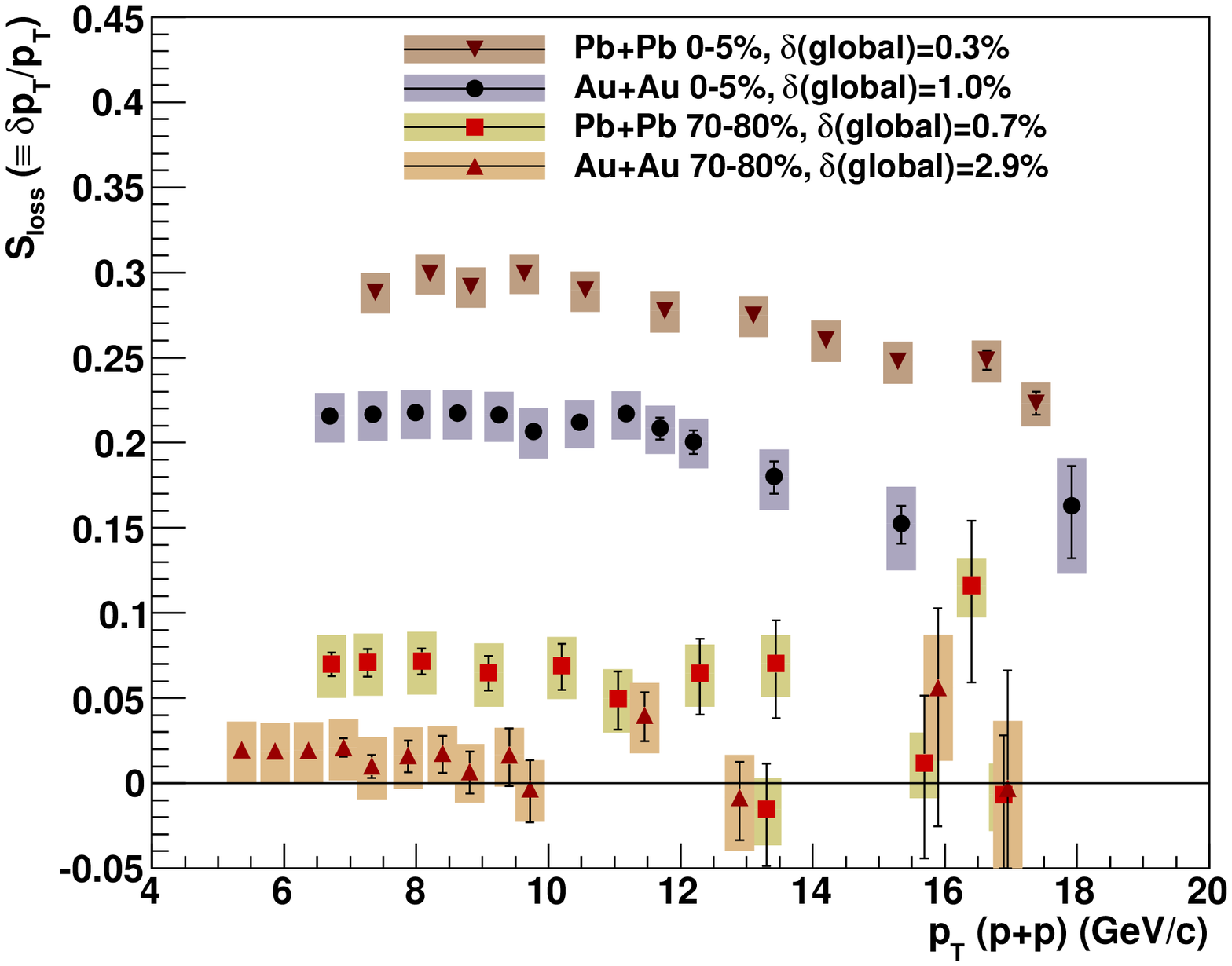}
  \caption{(Color online) 
  Comparison of average fractional
  momentum loss, as defined in the text, between the \sqsn = 200\,GeV
  \auau collisions (\piz, current paper) and \sqsn = 2.76\,TeV 
  \pbpb
  collisions (ALICE, charged hadrons~\cite{alice2011}).  The centrality
  selections are the same.
  $\delta$(global) stands for the uncertainty coming from the
   uncertainties of \taa. The overall normalization error from the
   \pp measurement is 1.3\,\% for \auau data, and is not shown here.
  }  
  \label{fig:sloss_alice}
\end{figure}

The average fractional momentum loss 
(\sloss) of high \pt hadrons
has been of interest since it may reflect the average fractional energy loss
of the initial parton.
\sloss is defined as $\delta p_{T}/p_{T}$, where $\delta p_{T}$ is
the difference of the momentum in \pp collisions 
($p_{\rm T,pp}$) and
that in \auau collisions [$p_{\rm T,AuAu}$], and the \pt in the 
denominator
is $p_{\rm T,pp}$.
In the previous publication~\cite{ppg054}, the assumption was made that both
\auau and \pp spectra are comparable in shape and \raa vs \pt is flat or
slowly varying, since the data sample size was not large enough to
directly calculate the $\delta p_{T}$.
With these assumptions, the suppression of high \pt hadrons
could be phenomenologically interpreted as a fractional momentum loss 
$\delta p_{T} / p_{T}$ by fitting \auau spectra with,
$f(p_T) = A\times[p_T(1+\delta p_T/p_T)]^{-n}$,
where $A$ and $n$ were obtained from by fitting a power-law function to
\taa-scaled \pp cross section~\cite{ppg054}.

With larger statistics \pp and \auau data collected, it is
possible to directly calculate \sloss without any assumptions.
The calculation method is schematically depicted in Fig.~\ref{fig:slossmethod}.
First, the \piz cross section in \pp [$f(p_T)$] is scaled by
\taa corresponding to the centrality selection of the \auau data
[$g(p_T)$].  Second, the scaled \pp cross section
[$T_{\rm AA}f(p_T)$] is fit with a power-law function [$h(p_T)$]. 
Third, the scaled \pp point closest in yield to the
Au$+$Au point of interest [$p\prime_{T,pp}$] is
found using the fit to interpolate between
$T_{AA}$ scaled \pp data points.
The $\delta p_T$ is calculated as 
$p_{\rm T,pp} - p_{\rm T,AuAu}$. For obtaining \sloss, the
$\delta p_T$ is divided by the $p_{\rm T,pp}$.
The uncertainty of the \sloss is calculated by inversely converting
the quadratic sum of the uncertainties on the yields of \auau and \pp
points, by the \pp fit function. Statistical and type B systematic
uncertainties are individually calculated in the same way. Therefore,
the $p_T$ dependence of systematic uncertainties are propagated to
the \sloss values.

Figure~\ref{fig:sloss} shows the results for minimum bias
collisions and three different centralities.
The uncertainty coming from \taa, which is of type C, changes
with centrality selection as listed on the plot. 
The \pp normalization
error of 9.7\% is not shown here because it moves all the points
independent of \pt or centrality.
Because \sloss is plotted as a function of \pt in \pp
collisions, the \pt points in successive centrality bins in \auau
are shifted as the momentum loss of hadrons varies.
An interesting feature of
the central collision data is that while $\delta p_T/p_T$ is constant
up to at least 10\,\gevc, at higher \pt it slowly decreases,
consistent with the slow rise of \raa.
If one assumes that the fragmentation function of the parton after
energy loss is unchanged,
the fractional momentum loss can be interpreted as the average fractional
energy loss $\langle \varepsilon \rangle = \langle \Delta E/E \rangle$
of the initial parton. 
This $\langle \varepsilon \rangle$ can then be compared to the 
trends predicted in~\cite{horowitz2011}.
In this particular model (see Figure 4 in~\cite{horowitz2011}), the
collisional energy loss appears to be somewhat overestimated,
particularly below 10\,\gevc, but at higher \pt the observed trend in 
$\delta p_T/p_T$ is reproduced quite well.

Figure~\ref{fig:phenixalicecomp} showed that the
\raa in the same centrality at RHIC and LHC show very similar $p_T$
dependence even though the collision systems and center-of-mass energies
are vastly different.  Figure~\ref{fig:sloss_alice} shows comparisons of
\sloss.   Note that the \sloss obtained
from the ALICE charged hadron measurement is $\sim$30\% higher than
that from the PHENIX \piz measurement. This is reasonable considering
the fact that the powers ($n$) in the power-law fit to the \pt spectra
are different between the two systems; the power of the PHENIX \pp \piz{s}
at $\sqrt{s}=200$\,\gevc is about 8, while that of the ALICE \pp charged
hadrons is about 6.


\subsection{Model calculations, transport coefficient}
\label{subsec:models}

\begin{figure}[htbp]
  \centering
  \includegraphics[width=1.0\linewidth]{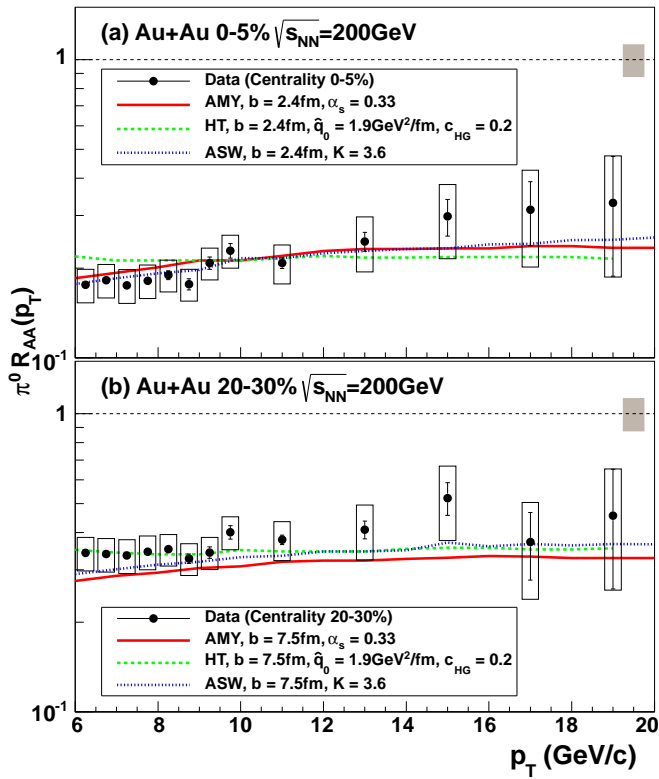}
  \caption{(Color online)
  (a) The \piz \raa 
  as a function of \pt at centrality 0--5\,\%.  The solid (red), 
  dashed (green), and dotted (blue) curves are the expectations of
  AMY~\cite{amy2001,amy2002}, HT~\cite{wang2001} and
  ASW~\cite{asw2003} models, respectively. 
  (b) The \piz \raa as a
  function of \pt at centrality 20--30\,\%.
  The theoretical curves in both panels are obtained
  from~\cite{bass2009}. The gray boxes around 1
  show global uncertainties and are of type C. 
}
  \label{fig:3dhydro_raa}
\end{figure}

In this section,
\raa is compared to four different parton energy loss models,
following the method described in~\cite{bass2009}.  All four models
are incorporated into the same three-dimensional relativistic hydrodynamic
calculation with an initial thermalization time $\tau_0$= 0.6\,fm/$c$
and describe the observed elliptic flow, pseudorapidity
distributions, and particle spectra at low \pt.
The Arnold-Moore-Yaffe
formalism (AMY~\cite{amy2001,amy2002}) incorporates radiative and
collisional energy loss processes in an extended medium in equilibrium
at high temperature, i.e. small coupling constant $g$, where
$\alpha_S=\frac{g^2}{4\pi}$.  In this approximation, a hierarchy of
scales of successively higher powers of the coupling constant can be
identified, and it becomes possible to construct an effective theory of
soft modes by summing contributions from hard loops into effective
propagators and vertices.
The higher-twist approach (HT~\cite{wang2001})
is based on the medium-enhanced higher-twist corrections to the total
cross section in deep inelastic scattering (DIS) off large
nuclei~\cite{qiu1991}.  HT incorporates only radiative corrections, but
it can
directly calculate the medium-modified fragmentation function.
The Armesto-Salgado-Wiedemann approach (ASW~\cite{asw2003}), which is
equivalent to the well known BDMPS-Z 
approach~\cite{Baier:1996kr, Zakharov:1997uu}, includes only
radiative processes in a medium where the mean free path of the parton
is much larger than the color-screening length.

The crucial parameter in all these models is the transport coefficient
$\hat{q}$ defined as

\begin{equation}
\hat{q} = \frac{\mu^2}{\lambda} [{\rm GeV^2/fm}]
\label{eq:qhat}
\end{equation}

\noindent
where $\mu^2$ is the average squared transverse momentum transferred
from the medium to the parton per collision and $\lambda$ is the mean
free path of the partons.  In AMY $\hat{q}$ is directly related to the
temperature, while in HT it is related to the local entropy density
$s$ ($\propto T^3$) and in ASW it is related to the energy density
$\varepsilon$.

Figure~\ref{fig:3dhydro_raa} compares the measured \raa 
at two centralities with calculations using
the energy loss models described above, 
incorporated into the same hydrodynamic
evolution~\cite{bass2009}.
In these models, the value of $\hat{q}$ is fixed
such as to reproduce the measured \raa in 0--5\% centrality collisions.
(See~\cite{bass2009} for the definitions of the  parameters 
$c_{{\rm HG}}$, and $K$,
which can be converted to $\hat{q_{0}}$.)
The values of $\hat{q_{0}}$ for gluons (defined as the value of $\hat{q}$ at
$\tau = 0.6$~fm/$c$ required to describe \raa) differ by a factor of
five: $\hat{q_{0}}$ is 4.1, 4.3 and
18.5~GeV$^{2}/$fm in AMY, HT and ASW, respectively.
The HT formalism was originally developed for deep inelastic scattering
off a large nucleus, and hence it has become customary to quote the value of
$\hat{q_{0}}$ for a quark~\cite{Majumder:2009cf}, and gives the value
$\hat{q_{0}}$=1.9~GeV$^{2}/$fm as seen in Fig.~\ref{fig:3dhydro_raa}.
Despite the large differences in the values of $\hat{q}$, all models describe
both the \pt-dependence and the centrality dependence of
\raa quite well.  Additional experimental
constraints are needed to differentiate between the models,
for instance, restricting the average path length \mean{L}
the parton traverses in the medium, which can be achieved not only in
two-particle correlation measurements~\cite{ppg090} but also by studying
\raadf of single particles.

\subsection{Nuclear modification factor vs event plane}

\begin{figure*}[htbp]
  \centering
  \includegraphics[width=0.99\linewidth]{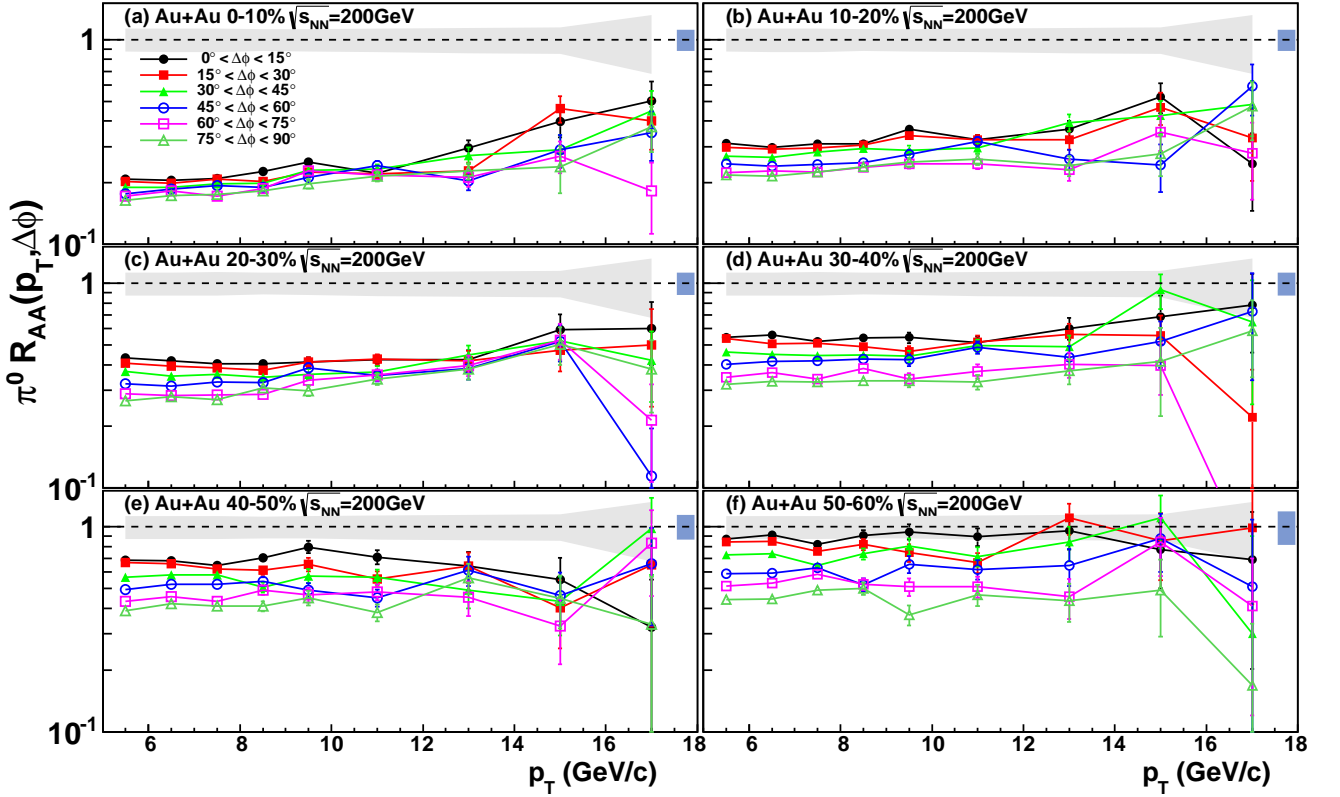}
  \caption{(Color online) \raadf as a function of $p_{T}$
  for the first six, 10\%-wide centrality classes.  Each of the six
  curves represent a 15$^{\circ}$-wide bin in azimuth, starting from
  $\phi=0^{\circ}$ (in-plane) up to $\phi=90^{\circ}$ (out of plane).
  The shaded (gray) band around 1 is the systematic uncertainty of 
  the normalizing $\phi$-integrated \raa. The shaded (blue) boxes 
  around 1 show global uncertainties and are of type C.
}
  \label{fig:RaaAllPhi}
\end{figure*}

The overlap region of the
colliding nuclei is not azimuthally isotropic, and neither is the medium that
is formed in the collision.  To first approximation (homogeneous
density distribution of nucleons) the overlap region is elliptical,
with the short axis being in the reaction plane.  As a consequence, the 
average path length the hard scattered parton traverses in the medium, 
losing energy in the process, varies with the azimuthal angle 
\df, defined experimentally as the relative azimuthal angle between
the emerging hadron and the measured event plane.
Measuring \raa
as a function of \df provides additional constraints on the average 
in-medium path length~\cite{ppg054,ppg092,ppg110}, 
therefore, a more stringent test of energy loss
models than the $\phi$-integrated \raa alone.

\begin{figure}[htbp]
  \centering
  \includegraphics[width=1.0\linewidth]{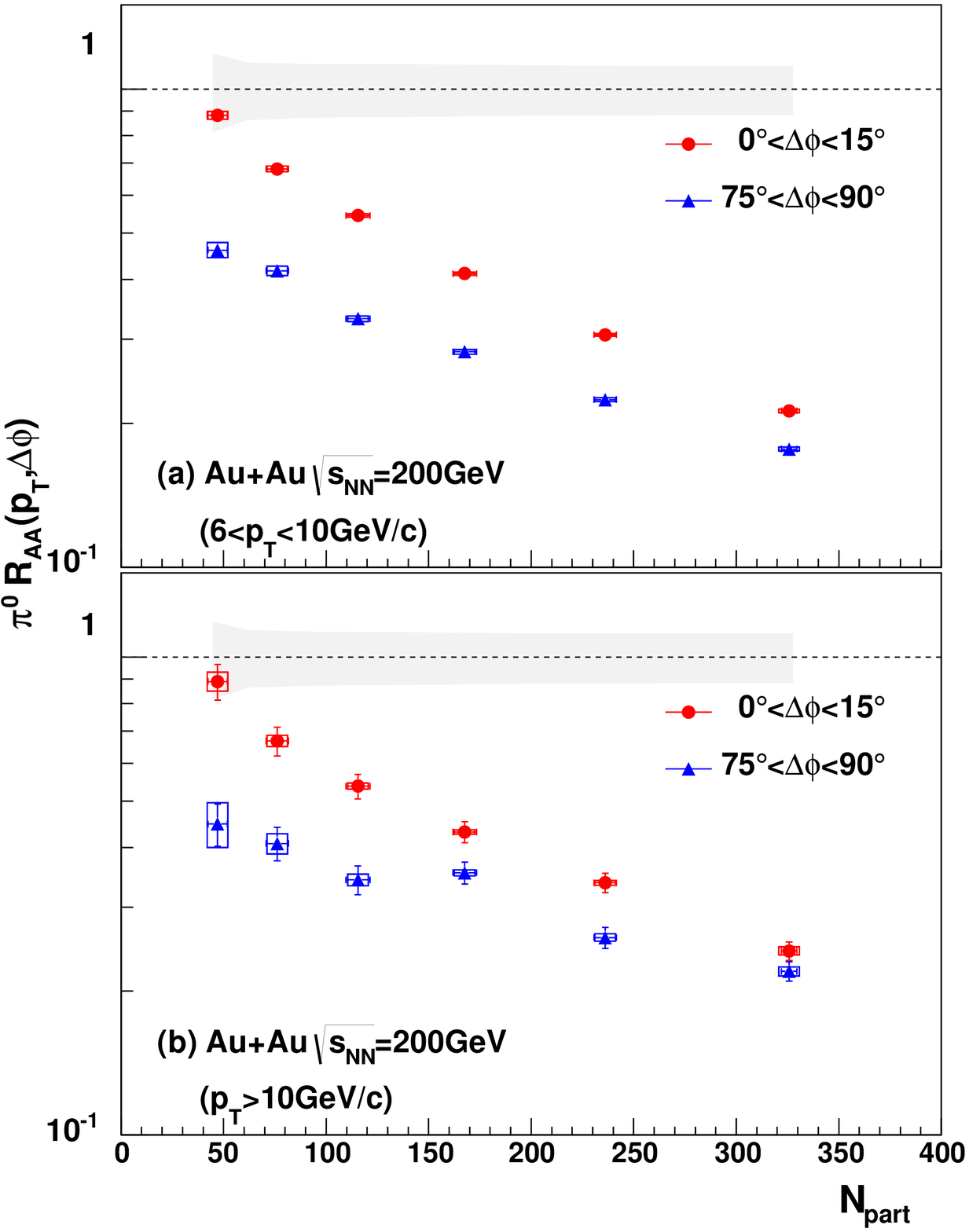}
  \caption{(Color online) Centrality dependence (expressed in terms of
  \Npart) of the \piz \raadf in-plane closed (red) circles
  and out-of-plane closed (blue) triangles,
  averaged (a) in the 6-10\,\gevc transverse momentum region and
  (b) above 10\,\gevc.
  Open boxes are systematic uncertainties on \raadf.
  } 
  \label{fig:intpi0raaphi}
\end{figure}

Figure~\ref{fig:RaaAllPhi} shows the differential nuclear modification
factor \raadf for six bins in azimuth and six centralities.
The participant eccentricities in Table~\ref{tab:glauber}
indicate the difference between the two extremes, in-plane and
out-of-plane.
In the most central collisions [panel (a)] the average path lengths
in-plane and out-of-plane are almost identical, therefore, the \raadf
curves almost completely overlap.  As one moves to more peripheral
collisions, the eccentricity of the overlap region increases and the
six curves start to split up showing the expected ordering:
suppression is always largest out-of-plane and smallest in-plane.
A simple calculation using the participant eccentricity 
(see Table~\ref{tab:glauber}) shows that the in-plane path length changes 
from 6.1\,fm to 3.4\,fm when 0--10\% and 50--60\% centralities are
compared, while the out-of-plane path length changes from 6.7\,fm to
5.9\,fm between the same two centralities.  As a consequence,
the out-of-plane \raadf changes much
less with centrality than the in-plane \raadf.  All these observations
are in full agreement with the findings in~\cite{ppg092}.

Figure~\ref{fig:intpi0raaphi} shows the evolution of \raadf
with centrality in-plane and out-of-plane at (a) moderate transverse
momenta (averaged in the 6--10\,\gevc \pt region) and (b) 
averaged over all available \pt above 10\,\gevc.
As expected, the difference between in-plane and out-of-plane
suppression increases with eccentricity (decreasing \Npart), and the 
actual values converge toward each other as the centrality increases.  

\begin{figure*}[htbp]
  \centering
  \includegraphics[width=0.99\linewidth]{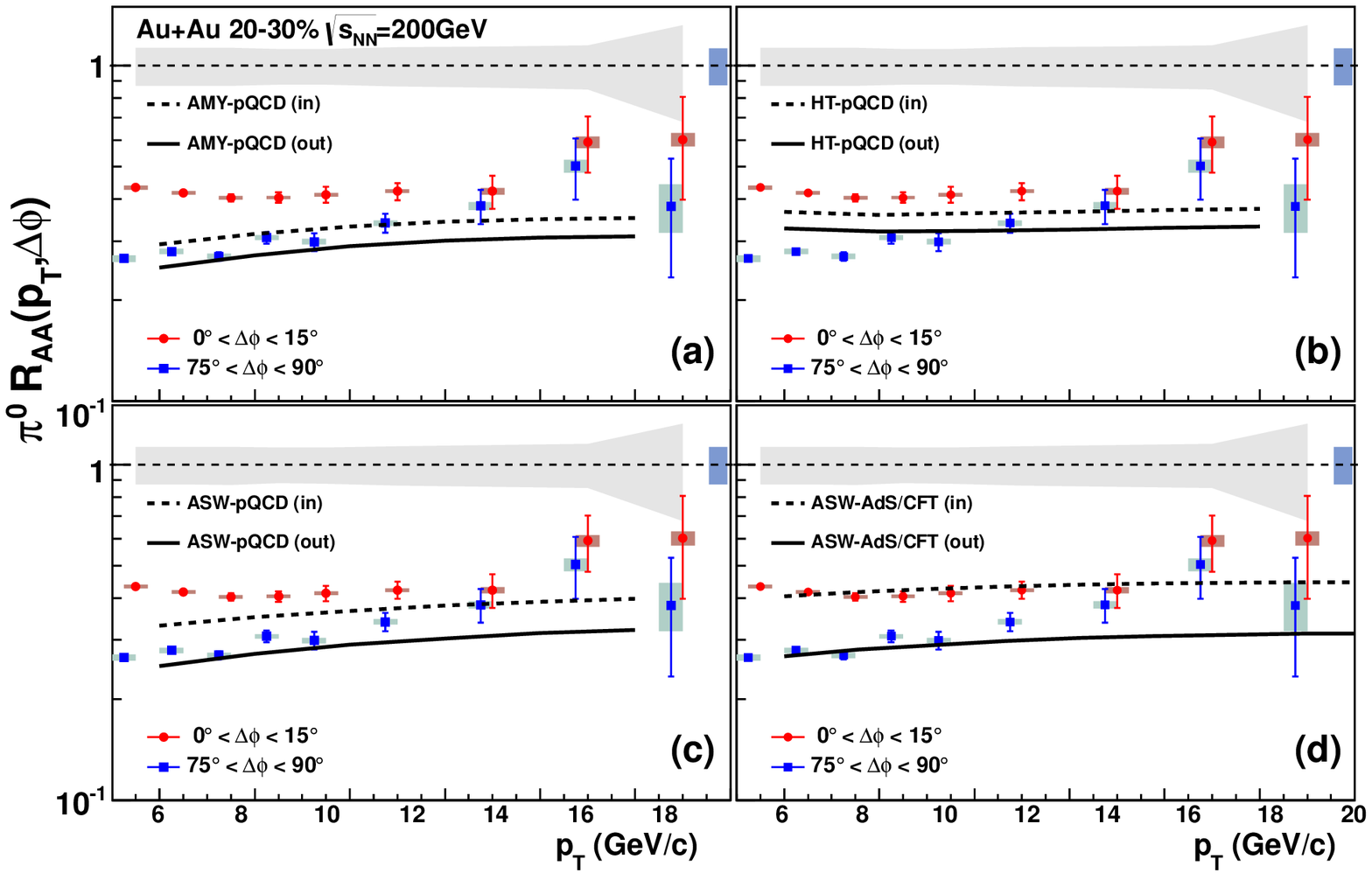}
  \caption{(Color online) 
  The data points are \raadf in 20--30\% centrality 
  as a function of \pt for 
  in-plane closed (red) circles and out-of-plane closed (blue) 
  triangles, compared
  to four model calculations (see text for description and references).  
  (a) pQCD-based AMY~\cite{amy2001, amy2002}, (b) HT~\cite{wang2001}, (c)
  pQCD-based ASW~\cite{asw2003}, (d) ASW using AdS/CFT
  correspondence~\cite{aswads2010}. The curves on panels
  (a)-(c) are taken from~\cite{bass2009}.
  The dashed and solid lines are the in-plane and out-of-plane
  predictions, respectively. The definition of the bands and
  boxes is the same as in Fig.~\protect\ref{fig:RaaAllPhi}.
}
  \label{fig:modelcomp}
\end{figure*}

Figure~\ref{fig:modelcomp} shows the comparisons of the models 
to the measured in-plane and out-of-plane \raa as a function of 
\pt for 20--30\% centrality.  The choice of the 20--30\% 
centrality interval is motivated by the availability of 
calculations for all the models shown.  Furthermore, this 
interval is a ``sweet spot" in determining the reaction plane 
(minimum uncertainty).  While statistical limitations of the 
reaction plane selected $R_{AA}$ vs $p_T$ do not prove that 
$R_{AA}$ rises with $p_T$, that rise is apparent from the 
reaction-plane-integrated measurement shown in 
Fig.~\ref{fig:RaaFitCentDep}.  The $\phi$ (i.e. pathlength) 
dependence is clear from the increasing in-plane vs out-of-plane 
difference in \raa vs centrality and the consistent ordering of 
the \raadf curves in Fig.~\ref{fig:RaaAllPhi}.

The brackets and bars on the data in Fig.~\ref{fig:modelcomp} are 
the statistical uncertainties of the in-plane and out-of-plane 
\raa. The shaded (gray) band around 1 corresponds to the 
systematic uncertainty of the average \piz \raa, while the shaded 
(blue) boxes at the right end of the \raa=1 lines show the 
uncertainty on \taa and are of type C. The shaded bands on the 
data points are the systematic uncertainty of the $dN/d\phi$ 
including the uncertainty from the event-plane resolution. Closed 
(red) circles and closed (blue) squares are the in-plane and 
out-of-plane \raa, respectively. Panel (a) shows the data overlaid 
with the AMY calculation~\cite{amy2001, amy2002}.  While the 
out-of-plane data are well described, the in-plane data are not, 
implying that the path-length dependence is too weak in this 
model.  The comparison with HT in panel (b) shows that this model 
fails to describe both the general trend and the in-plane vs. 
out-of-plane differences.  The ASW formalism [panel (c)] describes 
the out-of-plane suppression as well as AMY and shows a somewhat 
larger in-plane vs. out-of-plane difference, but is still 
inconsistent with the data.  It should be noted that in these 
three models the energy loss is proportional to $L^2$, where $L$ 
is the path length in the medium, and the quadratic dependence is 
characteristic when radiative energy loss is the dominant 
mechanism.

Finally, in panel (d) the data are compared to a model that 
invokes strong coupling in the medium via an AdS/CFT-inspired 
model.  The ASW-AdS/CFT formalism~\cite{aswads2010} incorporates 
the ASW treatment of hard processes, but for the soft processes 
assumes strong coupling.  Such a hybrid procedure was first 
suggested in~\cite{Liu:2006ug}. The virtual gluons radiated into 
the medium are governed by pQCD, but the interactions of those 
virtual gluons with the medium to bring them on shell is done by 
assuming that the transverse-momentum-squared is proportional to 
$L^2$ as given by an AdS/CFT 
calculation~\cite{aswads2010,Dominguez:2008vd}. This is in 
contrast to the weak-coupling expression for the 
transverse-momentum-squared, $\hat{q}L$. This results in an energy 
loss proportional to L$^3$ instead of L$^2$ as in the case of the 
pQCD-based models.  Panel (d) shows that the ASW-AdS/CFT model 
describes both the general shape and the absolute difference of 
the in-plane and out-of-plane data well.  The observation that 
models with path-length dependence of energy loss stronger than 
$L^2$ are in better agreement with the measurements is consistent 
with the findings in~\cite{ppg110}.

\section{Summary and Conclusions}

In summary, the large data set presented in this paper
made possible a measurement of the \piz invariant yield in \sqsn =
200\,GeV \auau collisions up to 20\,\gevc transverse momentum.
This has led to a precision measurement of
the $\eta$/\piz ratio in \auau collisions, which is constant as a function
of both centrality and
\pt, $\eta$/\piz = 0.45$\pm$0.01(stat)$\pm$0.04(syst)
and consistent with the values observed in \dau and \pp.
The large observed \piz suppression is fully consistent with
earlier findings, and a slow but significant rise of \raa vs \pt
with a slope of 0.0106$\pm^{0.0034}_{0.0029}$\,(\gevc)$^{-1}$
for central collisions is now observed for the first time at RHIC energies.
This has been an expectation of all pQCD-based parton energy loss models.
The large data set has also made possible the calculation of a
phenomenological $\Delta E/E$ energy loss.
The differential \raadf, testing the path-length dependence of energy
loss, is measured up to \pt of 20\,\gevc and is compared
to various energy loss calculations.  While all models considered
describe the $\phi$-integrated \raa adequately,
the pQCD based calculations where the energy loss depends on the
path length as $L^2$ fail to describe the differential \raadf. The data
require an energy loss with a power greater than 2, as given by models
in which the soft interactions with the medium are strongly coupled.

These findings are consistent with the conclusions of \cite{ppg110} in
which data on elliptic flow of high p$_T$ ($>$ 6 GeV)
$\pi^0$s is shown to be inconsistent with pQCD-based models. To explore
the  strong coupling regime, a comparison was made to the same 
ASW-AdS/CFT model
used in this work, as well as to a phenomenological
model~\cite{Jia:2010ee} in which the energy loss was proportional to
$L^3$, both of which were able to fit the data. Both the current 
measurement and~\cite{ppg110}
explore a region of high p$_T$ where the mechanism 
leading to an azimuthally anisotropic yield is 
parton energy loss rather than
hydrodynamical flow. It is increasingly difficult for purely 
pQCD-based models to explain these results and one is led to 
the tentative
conclusion that strong coupling plays an important role in parton
energy loss in the medium. At present, the best method to do the relevant
calculations is in an AdS/CFT framework. Similar conclusions are reached
when one looks at the behavior of heavy quarks~\cite{Adare:2010de}, where
higher quality data will soon be available. Recent preliminary results
on the suppression pattern seen at the LHC for p$_T >$ 6 GeV/$c$ are
strikingly similar to those seen at RHIC. In this paper, a phenomenological
calculation of fractional energy loss is given, which
indicates that the energy loss at LHC (ALICE data) is about 30\%
higher than at RHIC, and that the loss falls slightly with energy.
The dependence of these observables on momentum and
center-of-mass energy (presumably on energy density)
will be a crucial factor in untangling the underlying
mechanisms of parton energy loss. 

Recently, experiments at the LHC
have begun to examine the behavior of fully reconstructed jets,
which should give more easily interpretable information on this
phenomenon. Future work at both the LHC and at RHIC should bring data
on path length dependence of fully reconstructed jets, jet widths, and
heavy-quark jets which will add a wealth of new information. 
In addition, a more complete
understanding of the initial state is also needed both for the
initial configuration of hydrodynamical models, and as a calibration of
the hard probes that are used in these measurements.


\section*{ACKNOWLEDGMENTS}   

We thank the staff of the Collider-Accelerator and Physics
Departments at Brookhaven National Laboratory and the staff of
the other PHENIX participating institutions for their vital
contributions.  We acknowledge support from the 
Office of Nuclear Physics in the
Office of Science of the Department of Energy, the
National Science Foundation, Abilene Christian University
Research Council, Research Foundation of SUNY, and Dean of the
College of Arts and Sciences, Vanderbilt University (U.S.A),
Ministry of Education, Culture, Sports, Science, and Technology
and the Japan Society for the Promotion of Science (Japan),
Conselho Nacional de Desenvolvimento Cient\'{\i}fico e
Tecnol{\'o}gico and Funda\c c{\~a}o de Amparo {\`a} Pesquisa do
Estado de S{\~a}o Paulo (Brazil),
Natural Science Foundation of China (P.~R.~China),
Ministry of Education, Youth and Sports (Czech Republic),
Centre National de la Recherche Scientifique, Commissariat
{\`a} l'{\'E}nergie Atomique, and Institut National de Physique
Nucl{\'e}aire et de Physique des Particules (France),
Bundesministerium f\"ur Bildung und Forschung, Deutscher
Akademischer Austausch Dienst, and Alexander von Humboldt Stiftung (Germany),
Hungarian National Science Fund, OTKA (Hungary), 
Department of Atomic Energy and Department of Science and Technology (India), 
Israel Science Foundation (Israel), 
National Research Foundation and WCU program of the 
Ministry Education Science and Technology (Korea),
Ministry of Education and Science, Russian Academy of Sciences,
Federal Agency of Atomic Energy (Russia),
VR and Wallenberg Foundation (Sweden), 
the U.S. Civilian Research and Development Foundation for the
Independent States of the Former Soviet Union, 
the US-Hungarian Fulbright Foundation for Educational Exchange,
and the US-Israel Binational Science Foundation.


\section*{APPENDIX}

Tables~\ref{tab:invyield1} and \ref{tab:invyield2} give values for the 
invariant Yields for neutral pions, as shown in Fig.~\ref{fig:invyield}.  
Tables~\ref{tab:raa1} and \ref{tab:raa2} give values of 
$R_{\rm AA}$ for neutral pions, as shown in Fig.~\ref{fig:run7raa}.  

\begingroup \squeezetable

\begin{table}[th]
\caption{\label{tab:invyield1}
Invariant yields of neutral pions as a function of $p_{T}$ at
$|y|<0.35$ in Au$+$Au collisions at $\sqrt{s_{NN}}$=200~GeV
for the very-most central 0--5\% centrality.
Syst.(B) refers to type-B systematic errors.
See Fig.~\protect\ref{fig:invyield}.}
\begin{ruledtabular}\begin{tabular}{ccccccc}
Cen- & $p_T$ & Inv. Yield
& Stat. & Fraction & Syst.(B) & Fraction \\
trality  &   &
& error & \%  & error & \% \\
\hline
\\
0--5\%
& 5.25 & 9.394$\times 10^{-5}$ & 8.7$\times 10^{-7}$ & 0.92 & 8.4$\times 10^{-6}$ & 8.9 \\ 
& 5.75 & 4.524$\times 10^{-5}$ & 5.2$\times 10^{-7}$ & 1.1 & 4.0$\times 10^{-6}$ & 8.9 \\ 
& 6.25 & 2.273$\times 10^{-5}$ & 3.2$\times 10^{-7}$ & 1.4 & 2.0$\times 10^{-6}$ & 8.9 \\ 
& 6.75 & 1.253$\times 10^{-5}$ & 2.2$\times 10^{-7}$ & 1.7 & 1.1$\times 10^{-6}$ & 8.9 \\ 
& 7.25 & 6.862$\times 10^{-6}$ & 1.5$\times 10^{-7}$ & 2.1 & 6.1$\times 10^{-7}$ & 8.9 \\ 
& 7.75 & 4.164$\times 10^{-6}$ & 1.1$\times 10^{-7}$ & 2.5 & 3.7$\times 10^{-7}$ & 8.9 \\ 
& 8.25 & 2.598$\times 10^{-6}$ & 7.8$\times 10^{-8}$ & 3.0 & 2.0$\times 10^{-7}$ & 7.6 \\ 
& 8.75 & 1.545$\times 10^{-6}$ & 5.8$\times 10^{-8}$ & 3.7 & 1.2$\times 10^{-7}$ & 7.6 \\ 
& 9.25 & 1.118$\times 10^{-6}$ & 4.5$\times 10^{-8}$ & 4.0 & 8.6$\times 10^{-8}$ & 7.7 \\ 
& 9.75 & 7.684$\times 10^{-7}$ & 3.5$\times 10^{-8}$ & 4.6 & 6.3$\times 10^{-8}$ & 8.2 \\ 
& 11 & 2.837$\times 10^{-7}$ & 9.6$\times 10^{-9}$ & 3.4 & 3.1$\times 10^{-8}$ & 11 \\ 
& 13 & 8.685$\times 10^{-8}$ & 4.9$\times 10^{-9}$ & 5.7 & 1.5$\times 10^{-8}$ & 18 \\ 
& 15 & 2.659$\times 10^{-8}$ & 3.0$\times 10^{-9}$ & 11 & 6.7$\times 10^{-9}$ & 25 \\ 
& 17 & 9.547$\times 10^{-9}$ & 1.9$\times 10^{-9}$ & 20 & 3.1$\times 10^{-9}$ & 33 \\ 
& 19 & 4.450$\times 10^{-9}$ & 1.7$\times 10^{-9}$ & 38 & 1.8$\times 10^{-9}$ & 41 \\ 
\end{tabular}\end{ruledtabular}
\end{table}


\begin{table*}[th]
\caption{\label{tab:invyield2}
Invariant yields of neutral pions as a function of $p_{T}$ at 
$|y|<0.35$ in Au$+$Au collisions at $\sqrt{s_{NN}}$=200~GeV
for the indicated centrality ranges, including minimum bias (0--93\%).
Syst.(B) refers to type-B systematic errors.
See Fig.~\protect\ref{fig:invyield}.}
\begin{ruledtabular}\begin{tabular}{cccccccccccccccc}
Cen- & $p_T$ & Inv. Yield & Stat. & Fraction & Syst.(B) & Fraction && 
Cen- & $p_T$ & Inv. Yield & Stat. & Fraction & Syst.(B) & Fraction \\
trality  &   &  & error & \%  & error & \% & & 
trality  &   &  & error & \%  & error & \% \\
\hline  
0--10\%& 5.25 & 8.969$\times 10^{-5}$ & 5.6$\times 10^{-7}$ & 0.63 & 8.0$\times 10^{-6}$ & 
8.9  & & 	50--60\%& 5.25 & 2.147$\times 10^{-5}$ & 1.8$\times 10^{-7}$ & 0.84 & 
1.9$\times 10^{-6}$ & 8.9 \\
& 5.75 & 4.309$\times 10^{-5}$ & 3.4$\times 10^{-7}$ & 0.79 & 3.8$\times 10^{-6}$ & 8.9  & & 	
& 5.75 & 1.007$\times 10^{-5}$ & 1.2$\times 10^{-7}$ & 1.1 & 9.0$\times 10^{-7}$ & 8.9 \\
& 6.25 & 2.193$\times 10^{-5}$ & 2.2$\times 10^{-7}$ & 0.99 & 2.0$\times 10^{-6}$ & 8.9  & & 	
& 6.25 & 5.253$\times 10^{-6}$ & 7.9$\times 10^{-8}$ & 1.5 & 4.7$\times 10^{-7}$ & 8.9 \\
& 6.75 & 1.190$\times 10^{-5}$ & 1.4$\times 10^{-7}$ & 1.2 & 1.1$\times 10^{-6}$ & 8.9  & & 	
& 6.75 & 2.785$\times 10^{-6}$ & 5.4$\times 10^{-8}$ & 1.9 & 2.5$\times 10^{-7}$ & 8.9 \\
& 7.25 & 6.738$\times 10^{-6}$ & 1.0$\times 10^{-7}$ & 1.5 & 6.0$\times 10^{-7}$ & 8.9  & & 	
& 7.25 & 1.535$\times 10^{-6}$ & 3.9$\times 10^{-8}$ & 2.5 & 1.4$\times 10^{-7}$ & 8.9 \\
& 7.75 & 4.063$\times 10^{-6}$ & 7.2$\times 10^{-8}$ & 1.8 & 3.6$\times 10^{-7}$ & 8.9  & & 	
& 7.75 & 9.018$\times 10^{-7}$ & 2.8$\times 10^{-8}$ & 3.2 & 8.0$\times 10^{-8}$ & 8.9 \\
& 8.25 & 2.457$\times 10^{-6}$ & 5.3$\times 10^{-8}$ & 2.2 & 1.9$\times 10^{-7}$ & 7.6  & & 	
& 8.25 & 5.628$\times 10^{-7}$ & 2.2$\times 10^{-8}$ & 3.9 & 4.3$\times 10^{-8}$ & 7.6 \\
& 8.75 & 1.551$\times 10^{-6}$ & 4.0$\times 10^{-8}$ & 2.6 & 1.2$\times 10^{-7}$ & 7.6  & & 	
& 8.75 & 3.352$\times 10^{-7}$ & 1.6$\times 10^{-8}$ & 4.9 & 2.6$\times 10^{-8}$ & 7.6 \\
& 9.25 & 1.085$\times 10^{-6}$ & 3.1$\times 10^{-8}$ & 2.9 & 8.4$\times 10^{-8}$ & 7.7  & & 	
& 9.25 & 2.041$\times 10^{-7}$ & 1.3$\times 10^{-8}$ & 6.2 & 1.6$\times 10^{-8}$ & 7.7 \\
& 9.75 & 6.798$\times 10^{-7}$ & 2.4$\times 10^{-8}$ & 3.5 & 5.6$\times 10^{-8}$ & 8.2  & & 	
& 9.75 & 1.459$\times 10^{-7}$ & 1.0$\times 10^{-8}$ & 6.8 & 1.2$\times 10^{-8}$ & 8.2 \\
& 11 & 2.767$\times 10^{-7}$ & 6.7$\times 10^{-9}$ & 2.4 & 3.1$\times 10^{-8}$ & 11  & & 	
& 11 & 5.270$\times 10^{-8}$ & 2.8$\times 10^{-9}$ & 5.3 & 5.8$\times 10^{-9}$ & 11 \\
& 13 & 7.651$\times 10^{-8}$ & 3.3$\times 10^{-9}$ & 4.3 & 1.4$\times 10^{-8}$ & 18  & & 	
& 13 & 1.563$\times 10^{-8}$ & 1.7$\times 10^{-9}$ & 11 & 2.8$\times 10^{-9}$ & 18 \\
& 15 & 2.603$\times 10^{-8}$ & 2.0$\times 10^{-9}$ & 7.9 & 6.6$\times 10^{-9}$ & 25  & & 	
& 15 & 4.387$\times 10^{-9}$ & 7.8$\times 10^{-10}$ & 18 & 1.1$\times 10^{-9}$ & 25 \\
& 17 & 1.031$\times 10^{-8}$ & 1.4$\times 10^{-9}$ & 14 & 3.4$\times 10^{-9}$ & 33  & & 	
& 17 & 9.288$\times 10^{-10}$ & 4.2$\times 10^{-10}$ & 45 & 3.1$\times 10^{-10}$ & 33 \\
& 19 & 3.194$\times 10^{-9}$ & 1.0$\times 10^{-9}$ & 32 & 1.3$\times 10^{-9}$ & 41  & &   	
& 19 & 3.030$\times 10^{-10}$ & 3.0$\times 10^{-10}$ & 100 & 1.2$\times 10^{-10}$ & 41 \\ 
\\
10--20\%& 5.25 & 8.053$\times 10^{-5}$ & 4.6$\times 10^{-7}$ & 0.57 & 7.2$\times 10^{-6}$ & 
8.9  & & 	60--70\%& 5.25 & 1.155$\times 10^{-5}$ & 1.3$\times 10^{-7}$ & 1.1 & 
1.0$\times 10^{-6}$ & 8.9 \\
& 5.75 & 3.806$\times 10^{-5}$ & 2.8$\times 10^{-7}$ & 0.74 & 3.4$\times 10^{-6}$ & 8.9  & & 	
& 5.75 & 5.650$\times 10^{-6}$ & 8.4$\times 10^{-8}$ & 1.5 & 5.0$\times 10^{-7}$ & 8.9 \\
& 6.25 & 1.882$\times 10^{-5}$ & 1.8$\times 10^{-7}$ & 0.96 & 1.7$\times 10^{-6}$ & 8.9  & & 	
& 6.25 & 2.759$\times 10^{-6}$ & 5.7$\times 10^{-8}$ & 2.1 & 2.5$\times 10^{-7}$ & 8.9 \\
& 6.75 & 1.031$\times 10^{-5}$ & 1.2$\times 10^{-7}$ & 1.2 & 9.2$\times 10^{-7}$ & 8.9  & & 	
& 6.75 & 1.587$\times 10^{-6}$ & 4.0$\times 10^{-8}$ & 2.6 & 1.4$\times 10^{-7}$ & 8.9 \\
& 7.25 & 5.924$\times 10^{-6}$ & 8.7$\times 10^{-8}$ & 1.5 & 5.3$\times 10^{-7}$ & 8.9  & & 	
& 7.25 & 8.197$\times 10^{-7}$ & 2.8$\times 10^{-8}$ & 3.4 & 7.3$\times 10^{-8}$ & 8.9 \\
& 7.75 & 3.469$\times 10^{-6}$ & 6.2$\times 10^{-8}$ & 1.8 & 3.1$\times 10^{-7}$ & 8.9  & & 	
& 7.75 & 4.848$\times 10^{-7}$ & 2.1$\times 10^{-8}$ & 4.3 & 4.3$\times 10^{-8}$ & 8.9 \\
& 8.25 & 2.161$\times 10^{-6}$ & 4.7$\times 10^{-8}$ & 2.2 & 1.7$\times 10^{-7}$ & 7.6  & & 	
& 8.25 & 2.964$\times 10^{-7}$ & 1.6$\times 10^{-8}$ & 5.2 & 2.3$\times 10^{-8}$ & 7.6 \\
& 8.75 & 1.345$\times 10^{-6}$ & 3.5$\times 10^{-8}$ & 2.6 & 1.0$\times 10^{-7}$ & 7.6  & & 	
& 8.75 & 1.863$\times 10^{-7}$ & 1.2$\times 10^{-8}$ & 6.5 & 1.4$\times 10^{-8}$ & 7.6 \\
& 9.25 & 8.844$\times 10^{-7}$ & 2.7$\times 10^{-8}$ & 3.1 & 6.8$\times 10^{-8}$ & 7.7  & & 	
& 9.25 & 1.011$\times 10^{-7}$ & 9.1$\times 10^{-9}$ & 9.0 & 7.8$\times 10^{-9}$ & 7.7 \\
& 9.75 & 5.838$\times 10^{-7}$ & 2.1$\times 10^{-8}$ & 3.6 & 4.8$\times 10^{-8}$ & 8.2  & & 	
& 9.75 & 8.904$\times 10^{-8}$ & 8.1$\times 10^{-9}$ & 9.1 & 7.3$\times 10^{-9}$ & 8.2 \\
& 11 & 2.300$\times 10^{-7}$ & 5.9$\times 10^{-9}$ & 2.6 & 2.5$\times 10^{-8}$ & 11  & & 	
& 11 & 2.715$\times 10^{-8}$ & 2.0$\times 10^{-9}$ & 7.3 & 3.0$\times 10^{-9}$ & 11 \\
& 13 & 6.141$\times 10^{-8}$ & 2.9$\times 10^{-9}$ & 4.7 & 1.1$\times 10^{-8}$ & 18  & & 	
& 13 & 9.210$\times 10^{-9}$ & 1.1$\times 10^{-9}$ & 11 & 1.6$\times 10^{-9}$ & 18 \\
& 15 & 1.971$\times 10^{-8}$ & 1.7$\times 10^{-9}$ & 8.8 & 5.0$\times 10^{-9}$ & 25  & & 	
& 15 & 3.129$\times 10^{-9}$ & 6.5$\times 10^{-10}$ & 21 & 7.9$\times 10^{-10}$ & 25 \\
& 17 & 6.953$\times 10^{-9}$ & 1.3$\times 10^{-9}$ & 18 & 2.3$\times 10^{-9}$ & 33  & & 	
& 17 & 2.029$\times 10^{-9}$ & 6.1$\times 10^{-10}$ & 30 & 6.7$\times 10^{-10}$ & 33 \\
& 19 & 2.490$\times 10^{-9}$ & 8.8$\times 10^{-10}$ & 35 & 1.0$\times 10^{-9}$ & 41  & &   	
& 19 & 3.015$\times 10^{-10}$ & 3.0$\times 10^{-10}$ & 100 & 1.2$\times 10^{-10}$ & 41 \\
\\
20--30\%& 5.25 & 6.693$\times 10^{-5}$ & 3.8$\times 10^{-7}$ & 0.57 & 6.0$\times 10^{-6}$ & 
8.9  & & 	70--80\%& 5.25 & 5.486$\times 10^{-6}$ & 8.7$\times 10^{-8}$ & 1.6 & 
4.9$\times 10^{-7}$ & 8.9 \\
& 5.75 & 3.084$\times 10^{-5}$ & 2.3$\times 10^{-7}$ & 0.76 & 2.8$\times 10^{-6}$ & 8.9  & & 	
& 5.75 & 2.651$\times 10^{-6}$ & 5.7$\times 10^{-8}$ & 2.2 & 2.4$\times 10^{-7}$ & 8.9 \\
& 6.25 & 1.563$\times 10^{-5}$ & 1.5$\times 10^{-7}$ & 0.98 & 1.4$\times 10^{-6}$ & 8.9  & & 	
& 6.25 & 1.330$\times 10^{-6}$ & 3.9$\times 10^{-8}$ & 2.9 & 1.2$\times 10^{-7}$ & 8.9 \\
& 6.75 & 8.231$\times 10^{-6}$ & 1.0$\times 10^{-7}$ & 1.3 & 7.3$\times 10^{-7}$ & 8.9  & & 	
& 6.75 & 6.962$\times 10^{-7}$ & 2.6$\times 10^{-8}$ & 3.8 & 6.2$\times 10^{-8}$ & 8.9 \\
& 7.25 & 4.649$\times 10^{-6}$ & 7.3$\times 10^{-8}$ & 1.6 & 4.1$\times 10^{-7}$ & 8.9  & & 	
& 7.25 & 4.293$\times 10^{-7}$ & 2.0$\times 10^{-8}$ & 4.7 & 3.8$\times 10^{-8}$ & 8.9 \\
& 7.75 & 2.797$\times 10^{-6}$ & 5.4$\times 10^{-8}$ & 1.9 & 2.5$\times 10^{-7}$ & 8.9  & & 	
& 7.75 & 2.404$\times 10^{-7}$ & 1.5$\times 10^{-8}$ & 6.4 & 2.1$\times 10^{-8}$ & 8.9 \\
& 8.25 & 1.705$\times 10^{-6}$ & 4.0$\times 10^{-8}$ & 2.3 & 1.3$\times 10^{-7}$ & 7.6  & & 	
& 8.25 & 1.424$\times 10^{-7}$ & 1.1$\times 10^{-8}$ & 7.6 & 1.1$\times 10^{-8}$ & 7.6 \\
& 8.75 & 1.012$\times 10^{-6}$ & 3.0$\times 10^{-8}$ & 2.9 & 7.7$\times 10^{-8}$ & 7.6  & & 	
& 8.75 & 9.797$\times 10^{-8}$ & 8.3$\times 10^{-9}$ & 8.4 & 7.5$\times 10^{-9}$ & 7.6 \\
& 9.25 & 6.519$\times 10^{-7}$ & 2.3$\times 10^{-8}$ & 3.5 & 5.0$\times 10^{-8}$ & 7.7  & & 	
& 9.25 & 5.629$\times 10^{-8}$ & 6.6$\times 10^{-9}$ & 12 & 4.3$\times 10^{-9}$ & 7.7 \\
& 9.75 & 4.762$\times 10^{-7}$ & 1.8$\times 10^{-8}$ & 3.9 & 3.9$\times 10^{-8}$ & 8.2  & & 	
& 9.75 & 4.044$\times 10^{-8}$ & 5.0$\times 10^{-9}$ & 12 & 3.3$\times 10^{-9}$ & 8.2 \\
& 11 & 1.826$\times 10^{-7}$ & 5.2$\times 10^{-9}$ & 2.9 & 2.0$\times 10^{-8}$ & 11  & & 	
& 11 & 1.212$\times 10^{-8}$ & 1.2$\times 10^{-9}$ & 10 & 1.3$\times 10^{-9}$ & 11 \\
& 13 & 5.116$\times 10^{-8}$ & 2.6$\times 10^{-9}$ & 5.1 & 9.1$\times 10^{-9}$ & 18  & & 	
& 13 & 4.448$\times 10^{-9}$ & 7.3$\times 10^{-10}$ & 16 & 7.9$\times 10^{-10}$ & 18 \\
& 15 & 1.646$\times 10^{-8}$ & 1.6$\times 10^{-9}$ & 9.5 & 4.1$\times 10^{-9}$ & 25  & & 	
& 15 & 6.762$\times 10^{-9}$ & 3.0$\times 10^{-10}$ & 45 & 1.7$\times 10^{-10}$ & 25 \\
& 17 & 3.999$\times 10^{-9}$ & 8.7$\times 10^{-10}$ & 22 & 1.3$\times 10^{-9}$ & 33  & & 	
& 17 & 3.663$\times 10^{-9}$ & 2.6$\times 10^{-10}$ & 71 & 1.2$\times 10^{-10}$ & 33 \\
& 19 & 2.166$\times 10^{-9}$ & 8.2$\times 10^{-10}$ & 38 & 8.8$\times 10^{-10}$ & 41  & &    	
& 19 & --- & --- & --- & --- & --- \\ 
\\
30--40\%& 5.25 & 4.960$\times 10^{-5}$ & 3.0$\times 10^{-7}$ & 0.61 & 4.4$\times 10^{-6}$ & 
8.9  & & 	80--93\%& 5.25 & 1.748$\times 10^{-6}$ & 4.3$\times 10^{-8}$ & 2.5 & 
1.6$\times 10^{-7}$ & 8.9 \\
& 5.75 & 2.294$\times 10^{-5}$ & 1.9$\times 10^{-7}$ & 0.83 & 2.0$\times 10^{-6}$ & 8.9  & & 	
& 5.75 & 8.505$\times 10^{-7}$ & 2.8$\times 10^{-8}$ & 3.3 & 7.6$\times 10^{-8}$ & 8.9 \\
& 6.25 & 1.194$\times 10^{-5}$ & 1.3$\times 10^{-7}$ & 1.1 & 1.1$\times 10^{-6}$ & 8.9  & & 	
& 6.25 & 4.344$\times 10^{-7}$ & 2.0$\times 10^{-8}$ & 4.5 & 3.9$\times 10^{-8}$ & 8.9 \\
& 6.75 & 6.265$\times 10^{-6}$ & 8.7$\times 10^{-8}$ & 1.4 & 5.6$\times 10^{-7}$ & 8.9  & & 	
& 6.75 & 2.451$\times 10^{-7}$ & 1.4$\times 10^{-8}$ & 5.5 & 2.2$\times 10^{-8}$ & 8.9 \\
& 7.25 & 3.449$\times 10^{-6}$ & 6.1$\times 10^{-8}$ & 1.8 & 3.1$\times 10^{-7}$ & 8.9  & & 	
& 7.25 & 1.460$\times 10^{-7}$ & 1.0$\times 10^{-8}$ & 7.0 & 1.3$\times 10^{-8}$ & 8.9 \\
& 7.75 & 2.119$\times 10^{-6}$ & 4.5$\times 10^{-8}$ & 2.1 & 1.9$\times 10^{-7}$ & 8.9  & & 	
& 7.75 & 7.727$\times 10^{-8}$ & 7.2$\times 10^{-9}$ & 9.4 & 6.9$\times 10^{-9}$ & 8.9 \\
& 8.25 & 1.266$\times 10^{-6}$ & 3.3$\times 10^{-8}$ & 2.6 & 9.7$\times 10^{-8}$ & 7.6  & & 	
& 8.25 & 4.874$\times 10^{-8}$ & 5.6$\times 10^{-9}$ & 12 & 3.7$\times 10^{-9}$ & 7.6 \\
& 8.75 & 7.920$\times 10^{-7}$ & 2.5$\times 10^{-8}$ & 3.2 & 6.1$\times 10^{-8}$ & 7.6  & & 	
& 8.75 & 2.374$\times 10^{-8}$ & 4.4$\times 10^{-9}$ & 19 & 1.8$\times 10^{-9}$ & 7.6 \\
& 9.25 & 4.733$\times 10^{-7}$ & 1.9$\times 10^{-8}$ & 4.0 & 3.7$\times 10^{-8}$ & 7.7  & & 	
& 9.25 & 1.599$\times 10^{-8}$ & 2.8$\times 10^{-9}$ & 18 & 1.2$\times 10^{-9}$ & 7.7 \\
& 9.75 & 3.024$\times 10^{-7}$ & 1.5$\times 10^{-8}$ & 4.8 & 2.5$\times 10^{-8}$ & 8.2  & & 	
& 9.75 & 8.472$\times 10^{-9}$ & 2.0$\times 10^{-9}$ & 24 & 7.0$\times 10^{-10}$ & 8.2 \\
& 11 & 1.297$\times 10^{-7}$ & 4.3$\times 10^{-9}$ & 3.3 & 1.4$\times 10^{-8}$ & 11  & & 	
& 11 & 5.041$\times 10^{-9}$ & 7.0$\times 10^{-10}$ & 14 & 5.6$\times 10^{-10}$ & 11 \\
& 13 & 3.537$\times 10^{-8}$ & 2.2$\times 10^{-9}$ & 6.1 & 6.3$\times 10^{-9}$ & 18  & & 	
& 13 & 9.230$\times 10^{-10}$ & 2.9$\times 10^{-10}$ & 32 & 1.6$\times 10^{-10}$ & 18 \\
& 15 & 1.117$\times 10^{-8}$ & 1.5$\times 10^{-9}$ & 13 & 2.8$\times 10^{-9}$ & 25  & & 	
& 15 & 1.038$\times 10^{-10}$ & 1.0$\times 10^{-10}$ & 100 & 2.6$\times 10^{-11}$ & 25 \\
& 17 & 3.195$\times 10^{-9}$ & 7.7$\times 10^{-10}$ & 24 & 1.1$\times 10^{-9}$ & 33  & & 	
& 17 & 1.407$\times 10^{-10}$ & 1.4$\times 10^{-10}$ & 100 & 4.6$\times 10^{-11}$ & 33 \\
& 19 & 1.531$\times 10^{-9}$ & 6.8$\times 10^{-10}$ & 45 & 6.2$\times 10^{-10}$ & 41  & &    	
& 19 & --- & --- & --- & --- & --- \\ 
\\
40--50\%& 5.25 & 3.395$\times 10^{-5}$ & 2.3$\times 10^{-7}$ & 0.69 & 3.0$\times 10^{-6}$ & 
8.9  & & 	0--93\%& 5.25 & 4.038$\times 10^{-5}$ & 1.0$\times 10^{-7}$ & 0.26 & 
3.6$\times 10^{-6}$ & 8.9 \\
& 5.75 & 1.612$\times 10^{-5}$ & 1.5$\times 10^{-7}$ & 0.93 & 1.4$\times 10^{-6}$ & 8.9  & & 	
& 5.75 & 1.880$\times 10^{-5}$ & 6.3$\times 10^{-8}$ & 0.33 & 1.7$\times 10^{-6}$ & 8.9 \\
& 6.25 & 8.369$\times 10^{-6}$ & 1.0$\times 10^{-7}$ & 1.2 & 7.5$\times 10^{-7}$ & 8.9  & & 	
& 6.25 & 9.464$\times 10^{-6}$ & 4.1$\times 10^{-8}$ & 0.43 & 8.4$\times 10^{-7}$ & 8.9 \\
& 6.75 & 4.444$\times 10^{-6}$ & 7.0$\times 10^{-8}$ & 1.6 & 4.0$\times 10^{-7}$ & 8.9  & & 	
& 6.75 & 5.047$\times 10^{-6}$ & 2.7$\times 10^{-8}$ & 0.54 & 4.5$\times 10^{-7}$ & 8.9 \\
& 7.25 & 2.397$\times 10^{-6}$ & 5.0$\times 10^{-8}$ & 2.1 & 2.1$\times 10^{-7}$ & 8.9  & & 	
& 7.25 & 2.821$\times 10^{-6}$ & 1.9$\times 10^{-8}$ & 0.68 & 2.5$\times 10^{-7}$ & 8.9 \\
& 7.75 & 1.492$\times 10^{-6}$ & 3.7$\times 10^{-8}$ & 2.5 & 1.3$\times 10^{-7}$ & 8.9  & & 	
& 7.75 & 1.688$\times 10^{-6}$ & 1.4$\times 10^{-8}$ & 0.83 & 1.5$\times 10^{-7}$ & 8.9 \\
& 8.25 & 8.663$\times 10^{-7}$ & 2.7$\times 10^{-8}$ & 3.1 & 6.6$\times 10^{-8}$ & 7.6  & & 	
& 8.25 & 1.022$\times 10^{-6}$ & 1.0$\times 10^{-8}$ & 1.0 & 7.8$\times 10^{-8}$ & 7.6 \\
& 8.75 & 5.578$\times 10^{-7}$ & 2.1$\times 10^{-8}$ & 3.7 & 4.3$\times 10^{-8}$ & 7.6  & & 	
& 8.75 & 6.311$\times 10^{-7}$ & 7.8$\times 10^{-9}$ & 1.2 & 4.8$\times 10^{-8}$ & 7.6 \\
& 9.25 & 3.434$\times 10^{-7}$ & 1.6$\times 10^{-8}$ & 4.5 & 2.6$\times 10^{-8}$ & 7.7  & & 	
& 9.25 & 4.075$\times 10^{-7}$ & 6.0$\times 10^{-9}$ & 1.5 & 3.1$\times 10^{-8}$ & 7.7 \\
& 9.75 & 2.377$\times 10^{-7}$ & 1.3$\times 10^{-8}$ & 5.3 & 2.0$\times 10^{-8}$ & 8.2  & & 	
& 9.75 & 2.744$\times 10^{-7}$ & 4.7$\times 10^{-9}$ & 1.7 & 2.3$\times 10^{-8}$ & 8.2 \\
& 11 & 8.341$\times 10^{-8}$ & 3.5$\times 10^{-9}$ & 4.2 & 9.2$\times 10^{-9}$ & 11  & & 	
& 11 & 1.073$\times 10^{-7}$ & 1.3$\times 10^{-9}$ & 1.2 & 1.2$\times 10^{-8}$ & 11 \\
& 13 & 2.352$\times 10^{-8}$ & 1.8$\times 10^{-9}$ & 7.7 & 4.2$\times 10^{-9}$ & 18  & & 	
& 13 & 2.968$\times 10^{-8}$ & 6.6$\times 10^{-10}$ & 2.2 & 5.3$\times 10^{-9}$ & 18 \\
& 15 & 4.544$\times 10^{-9}$ & 2.6$\times 10^{-9}$ & 56 & 1.1$\times 10^{-9}$ & 25  & & 	
& 15 & 9.454$\times 10^{-9}$ & 4.0$\times 10^{-10}$ & 4.2 & 2.4$\times 10^{-9}$ & 25 \\
& 17 & 2.233$\times 10^{-9}$ & 6.4$\times 10^{-10}$ & 29 & 7.3$\times 10^{-10}$ & 33  & & 	
& 17 & 3.208$\times 10^{-9}$ & 2.6$\times 10^{-10}$ & 8.1 & 1.1$\times 10^{-9}$ & 33 \\
& 19 & 1.517$\times 10^{-9}$ & 6.8$\times 10^{-10}$ & 45 & 6.2$\times 10^{-10}$ & 41  & & 	
& 19 & 1.224$\times 10^{-9}$ & 2.0$\times 10^{-10}$ & 16 & 5.0$\times 10^{-10}$ & 41 \\
\end{tabular}\end{ruledtabular}
\end{table*}

\begin{table*}[th]
\caption{\label{tab:raa1}
Nuclear modification factors, $R_{\rm AA}$ for neutral pions as a function 
of $p_{T}$ at $|y|<0.35$ 
in Au$+$Au collisions at $\sqrt{s_{NN}}$=200~GeV
for the indicated centrality ranges, including minimum bias (0--93\%).
Syst.(B) refers to type-B systematic errors.
The global systematic uncertainties (type C) are $p$$+$$p$
normalization (9.7$\%$) and Off-vertex (1.5$\%$).  
See Fig.~\ref{fig:run7raa}.
}
\begin{ruledtabular}\begin{tabular}{cccccccccccccccc}
Cen- & $p_T$ & $R_{\rm AA}$ & Stat. & Fraction & Syst.(B) & Fraction & &
Cen- & $p_T$ & $R_{\rm AA}$ & Stat. & Fraction & Syst.(B) & Fraction \\
trality  &   &  & error & \%  & error & \% & &
trality  &   &  & error & \%  & error & \% \\ 
\hline 
0--10\%&5.25 & 0.1859 & 0.0014 & 0.74 & 0.024 & 13  & & 	50--60\%&5.25 & 0.6691 & 0.0062 & 0.92 & 0.086 & 13 \\
&5.75 & 0.1855 & 0.0018 & 0.95 & 0.024 & 13  & & 	&5.75 & 0.6517 & 0.0082 & 1.3 & 0.084 & 13 \\
&6.25 & 0.1882 & 0.0023 & 1.2 & 0.024 & 13  & & 	&6.25 & 0.6776 & 0.011 & 1.7 & 0.087 & 13 \\
&6.75 & 0.1922 & 0.0029 & 1.5 & 0.025 & 13  & & 	&6.75 & 0.6765 & 0.015 & 2.2 & 0.087 & 13 \\
&7.25 & 0.1908 & 0.0036 & 1.9 & 0.025 & 13  & & 	&7.25 & 0.6533 & 0.018 & 2.8 & 0.084 & 13 \\
&7.75 & 0.1967 & 0.0045 & 2.3 & 0.025 & 13  & & 	&7.75 & 0.6562 & 0.023 & 3.5 & 0.085 & 13 \\
&8.25 & 0.1990 & 0.0056 & 2.8 & 0.024 & 12  & & 	&8.25 & 0.6851 & 0.030 & 4.3 & 0.083 & 12 \\
&8.75 & 0.1970 & 0.0066 & 3.4 & 0.024 & 12  & & 	&8.75 & 0.6399 & 0.034 & 5.4 & 0.078 & 12 \\
&9.25 & 0.2241 & 0.0089 & 4.0 & 0.028 & 12  & & 	&9.25 & 0.6339 & 0.043 & 6.8 & 0.078 & 12 \\
&9.75 & 0.2250 & 0.011 & 4.9 & 0.028 & 13  & & 	&9.75 & 0.7255 & 0.055 & 7.6 & 0.092 & 13 \\
&11 & 0.2253 & 0.0077 & 3.4 & 0.034 & 15  & & 	&11 & 0.6448 & 0.038 & 5.9 & 0.096 & 15 \\
&13 & 0.2403 & 0.015 & 6.3 & 0.050 & 21  & & 	&13 & 0.7378 & 0.085 & 12 & 0.15 & 21 \\
&15 & 0.3244 & 0.037 & 11 & 0.091 & 28  & & 	&15 & 0.8217 & 0.16 & 19 & 0.23 & 28 \\
&17 & 0.3763 & 0.072 & 19 & 0.13 & 36  & & 	&17 & 0.5097 & 0.24 & 47 & 0.18 & 36 \\
&19 & 0.2639 & 0.10 & 38 & 0.12 & 44  & &  	&19 & 0.3762 & 0.38 & 100 & 0.16 & 44 \\ 
\\ %
10-20\%&5.25 & 0.2630 & 0.0018 & 0.69 & 0.034 & 13  & & 	60-70\%&5.25 & 0.7726 & 0.0091 & 1.2 & 0.099 & 13 \\
&5.75 & 0.2581 & 0.0023 & 0.91 & 0.033 & 13  & & 	&5.75 & 0.7851 & 0.012 & 1.6 & 0.10 & 13 \\
&6.25 & 0.2545 & 0.0030 & 1.2 & 0.033 & 13  & & 	&6.25 & 0.7642 & 0.017 & 2.2 & 0.098 & 13 \\
&6.75 & 0.2625 & 0.0040 & 1.5 & 0.034 & 13  & & 	&6.75 & 0.8278 & 0.022 & 2.7 & 0.11 & 13 \\
&7.25 & 0.2643 & 0.0050 & 1.9 & 0.034 & 13  & & 	&7.25 & 0.7490 & 0.027 & 3.6 & 0.097 & 13 \\
&7.75 & 0.2646 & 0.0061 & 2.3 & 0.034 & 13  & & 	&7.75 & 0.7576 & 0.034 & 4.5 & 0.098 & 13 \\
&8.25 & 0.2757 & 0.0078 & 2.8 & 0.034 & 12  & & 	&8.25 & 0.7747 & 0.043 & 5.5 & 0.094 & 12 \\
&8.75 & 0.2691 & 0.0091 & 3.4 & 0.033 & 12  & & 	&8.75 & 0.7636 & 0.053 & 6.9 & 0.093 & 12 \\
&9.25 & 0.2879 & 0.012 & 4.1 & 0.035 & 12  & & 	&9.25 & 0.6745 & 0.064 & 9.4 & 0.083 & 12 \\
&9.75 & 0.3043 & 0.015 & 5.0 & 0.038 & 13  & & 	&9.75 & 0.9509 & 0.092 & 9.7 & 0.12 & 13 \\
&11 & 0.2950 & 0.010 & 3.5 & 0.044 & 15  & & 	&11 & 0.7135 & 0.055 & 7.7 & 0.11 & 15 \\
&13 & 0.3038 & 0.020 & 6.6 & 0.063 & 21  & & 	&13 & 0.9334 & 0.12 & 12 & 0.19 & 21 \\
&15 & 0.3870 & 0.046 & 12 & 0.11 & 28  & & 	&15 & 1.259 & 0.28 & 22 & 0.35 & 28 \\
&17 & 0.4000 & 0.089 & 22 & 0.14 & 36  & & 	&17 & 2.391 & 0.79 & 33 & 0.85 & 36 \\
&19 & 0.3240 & 0.13 & 41 & 0.14 & 44  & &  	&19 & 0.8040 & 0.82 & 100 & 0.35 & 44 \\
\\ %
20--30\%&5.25 & 0.3528 & 0.0024 & 0.69 & 0.045 & 13  & & 	70--80\%&5.25 & 0.8701 & 0.014 & 1.6 & 0.11 & 13 \\
&5.75 & 0.3377 & 0.0031 & 0.92 & 0.043 & 13  & & 	&5.75 & 0.8731 & 0.019 & 2.2 & 0.11 & 13 \\
&6.25 & 0.3412 & 0.0041 & 1.2 & 0.044 & 13  & & 	&6.25 & 0.8736 & 0.026 & 3.0 & 0.11 & 13 \\
&6.75 & 0.3382 & 0.0053 & 1.6 & 0.044 & 13  & & 	&6.75 & 0.8608 & 0.034 & 3.9 & 0.11 & 13 \\
&7.25 & 0.3347 & 0.0065 & 2.0 & 0.043 & 13  & & 	&7.25 & 0.9298 & 0.045 & 4.9 & 0.12 & 13 \\
&7.75 & 0.3444 & 0.0083 & 2.4 & 0.045 & 13  & & 	&7.75 & 0.8904 & 0.059 & 6.6 & 0.12 & 13 \\
&8.25 & 0.3511 & 0.010 & 3.0 & 0.043 & 12  & & 	&8.25 & 0.8823 & 0.069 & 7.8 & 0.11 & 12 \\
&8.75 & 0.3268 & 0.012 & 3.6 & 0.040 & 12  & & 	&8.75 & 0.9520 & 0.083 & 8.7 & 0.12 & 12 \\
&9.25 & 0.3425 & 0.015 & 4.4 & 0.042 & 12  & & 	&9.25 & 0.8898 & 0.11 & 12 & 0.11 & 12 \\
&9.75 & 0.4006 & 0.021 & 5.1 & 0.051 & 13  & & 	&9.75 & 1.024 & 0.13 & 13 & 0.13 & 13 \\
&11 & 0.3779 & 0.014 & 3.7 & 0.057 & 15  & & 	&11 & 0.7552 & 0.079 & 10 & 0.11 & 15 \\
&13 & 0.4085 & 0.028 & 6.9 & 0.085 & 21  & & 	&13 & 1.069 & 0.18 & 17 & 0.22 & 21 \\
&15 & 0.5216 & 0.065 & 12 & 0.15 & 28  & & 	&15 & 0.6448 & 0.29 & 45 & 0.18 & 28 \\
&17 & 0.3713 & 0.095 & 26 & 0.13 & 36  & & 	&17 & 1.023 & 0.74 & 72 & 0.37 & 36 \\
&19 & 0.4550 & 0.20 & 43 & 0.20 & 44  & &   	&19 & --- & --- & --- & --- & --- \\
\\ %
30--40\%&5.25 & 0.4408 & 0.0032 & 0.72 & 0.056 & 13  & & 	80--93\%&5.25 & 0.8298 & 0.021 & 2.5 & 0.11 & 13 \\
&5.75 & 0.4234 & 0.0041 & 0.98 & 0.054 & 13  & & 	&5.75 & 0.8385 & 0.028 & 3.4 & 0.11 & 13 \\
&6.25 & 0.4394 & 0.0056 & 1.3 & 0.056 & 13  & & 	&6.25 & 0.8537 & 0.039 & 4.6 & 0.11 & 13 \\
&6.75 & 0.4341 & 0.0072 & 1.7 & 0.056 & 13  & & 	&6.75 & 0.9070 & 0.051 & 5.6 & 0.12 & 13 \\
&7.25 & 0.4186 & 0.0089 & 2.1 & 0.054 & 13  & & 	&7.25 & 0.9464 & 0.067 & 7.1 & 0.12 & 13 \\
&7.75 & 0.4398 & 0.011 & 2.6 & 0.057 & 13  & & 	&7.75 & 0.8566 & 0.081 & 9.5 & 0.11 & 13 \\
&8.25 & 0.4397 & 0.014 & 3.2 & 0.053 & 12  & & 	&8.25 & 0.9038 & 0.11 & 12 & 0.11 & 12 \\
&8.75 & 0.4313 & 0.017 & 3.9 & 0.053 & 12  & & 	&8.75 & 0.6904 & 0.13 & 19 & 0.084 & 12 \\
&9.25 & 0.4193 & 0.020 & 4.9 & 0.052 & 12  & & 	&9.25 & 0.7566 & 0.14 & 18 & 0.093 & 12 \\
&9.75 & 0.4291 & 0.025 & 5.9 & 0.054 & 13  & & 	&9.75 & 0.6419 & 0.15 & 24 & 0.081 & 13 \\
&11 & 0.4526 & 0.019 & 4.1 & 0.068 & 15  & & 	&11 & 0.9396 & 0.13 & 14 & 0.14 & 15 \\
&13 & 0.4762 & 0.036 & 7.6 & 0.099 & 21  & & 	&13 & 0.6637 & 0.21 & 32 & 0.14 & 21 \\
&15 & 0.5969 & 0.092 & 15 & 0.17 & 28  & & 	&15 & 0.2963 & 0.30 & 100 & 0.083 & 28 \\
&17 & 0.5002 & 0.14 & 28 & 0.18 & 36  & & 	&17 & 1.176 & 1.2 & 100 & 0.42 & 36 \\
&19 & 0.5423 & 0.27 & 49 & 0.24 & 44  & &  	&19 & --- & --- & --- & --- & --- \\ 
\\ %
40--50\%&5.25 & 0.5422 & 0.0043 & 0.79 & 0.069 & 13  & & 	0--93\%&5.25 & 0.3105 & 0.0014 & 0.46 & 0.040 & 13 \\
&5.75 & 0.5346 & 0.0057 & 1.1 & 0.069 & 13  & & 	&5.75 & 0.3002 & 0.0019 & 0.62 & 0.038 & 13 \\
&6.25 & 0.5532 & 0.0078 & 1.4 & 0.071 & 13  & & 	&6.25 & 0.3013 & 0.0025 & 0.82 & 0.039 & 13 \\
&6.75 & 0.5531 & 0.010 & 1.8 & 0.071 & 13  & & 	&6.75 & 0.3025 & 0.0032 & 1.1 & 0.039 & 13 \\
&7.25 & 0.5227 & 0.012 & 2.4 & 0.067 & 13  & & 	&7.25 & 0.2962 & 0.0040 & 1.4 & 0.038 & 13 \\
&7.75 & 0.5562 & 0.016 & 2.9 & 0.072 & 13  & & 	&7.75 & 0.3031 & 0.0051 & 1.7 & 0.039 & 13 \\
&8.25 & 0.5404 & 0.020 & 3.6 & 0.066 & 12  & & 	&8.25 & 0.3070 & 0.0064 & 2.1 & 0.037 & 12 \\
&8.75 & 0.5457 & 0.024 & 4.3 & 0.067 & 12  & & 	&8.75 & 0.2973 & 0.0074 & 2.5 & 0.036 & 12 \\
&9.25 & 0.5466 & 0.029 & 5.3 & 0.067 & 12  & & 	&9.25 & 0.3123 & 0.0096 & 3.1 & 0.038 & 12 \\
&9.75 & 0.6059 & 0.038 & 6.3 & 0.077 & 13  & & 	&9.75 & 0.3368 & 0.013 & 3.8 & 0.043 & 13 \\
&11 & 0.5231 & 0.025 & 4.8 & 0.078 & 15  & & 	&11 & 0.3240 & 0.0088 & 2.7 & 0.048 & 15 \\
&13 & 0.5690 & 0.051 & 9.0 & 0.12 & 21  & & 	&13 & 0.3458 & 0.018 & 5.1 & 0.072 & 21 \\
&15 & 0.4362 & 0.25 & 57 & 0.12 & 28  & & 	&15 & 0.4370 & 0.040 & 9.2 & 0.12 & 28 \\
&17 & 0.6280 & 0.20 & 32 & 0.22 & 36  & & 	&17 & 0.4344 & 0.067 & 16 & 0.16 & 36 \\
&19 & 0.9655 & 0.48 & 49 & 0.42 & 44  & & 	&19 & 0.3750 & 0.10 & 27 & 0.16 & 44 \\
\end{tabular}\end{ruledtabular}
\end{table*}

\clearpage

\begin{table}[th]
\caption{\label{tab:raa2}
Nuclear modification factors, $R_{\rm AA}$ for neutral pions 
as a function of $p_{T}$ at $|y|<0.35$
in Au$+$Au collisions at $\sqrt{s_{NN}}$=200~GeV
for the very most central, 0--5\% collisions. 
Syst.(B) refers to type-B systematic errors.
The global systematic uncertainties (type C) are $p$$+$$p$
normalization (9.7$\%$) and Off-vertex (1.5$\%$).  See Fig.~\ref{fig:run7raa}.
}
\begin{ruledtabular}\begin{tabular}{ccccccc}
Cen- & $p_T$ & $R_{\rm AA}$
& Stat. & Fraction & Syst.(B) & Fraction \\
trality  &   &  
& error & \%  & error & \% \\ 
\hline
\\
0--5\%
& 5.25 & 0.1753 & 0.0018 & 1.0 & 0.022 & 13 \\ 
& 5.75 & 0.1753 & 0.0022 & 1.3 & 0.022 & 13 \\ 
& 6.25 & 0.1756 & 0.0028 & 1.6 & 0.023 & 13 \\ 
& 6.75 & 0.1823 & 0.0036 & 2.0 & 0.023 & 13 \\ 
& 7.25 & 0.1749 & 0.0043 & 2.4 & 0.023 & 13 \\ 
& 7.75 & 0.1815 & 0.0053 & 2.9 & 0.023 & 13 \\ 
& 8.25 & 0.1894 & 0.0067 & 3.5 & 0.023 & 12 \\ 
& 8.75 & 0.1766 & 0.0076 & 4.3 & 0.022 & 12 \\ 
& 9.25 & 0.2080 & 0.010 & 4.8 & 0.026 & 12 \\ 
& 9.75 & 0.2288 & 0.013 & 5.7 & 0.029 & 13 \\ 
& 11 & 0.2079 & 0.0087 & 4.2 & 0.031 & 15 \\ 
& 13 & 0.2455 & 0.018 & 7.3 & 0.051 & 21 \\ 
& 15 & 0.2982 & 0.042 & 14 & 0.083 & 28 \\ 
& 17 & 0.3137 & 0.076 & 24 & 0.11 & 36 \\ 
& 19 & 0.3308 & 0.14 & 43 & 0.14 & 44 \\ 
\end{tabular}\end{ruledtabular}
\end{table}

\endgroup



\end{document}